\documentclass{article}

\usepackage[backend=biber,sorting=none,citestyle=numeric-comp,giveninits=true,doi=false,date=year,url=false,backref=true,maxbibnames=5]{biblatex}
\renewbibmacro{in:}{}

\DeclareFieldFormat{eprint:arxiv}{arXiv: \href{http://arxiv.org/abs/#1}{#1}}
\DeclareFieldFormat{pages}{\mkfirstpage{#1}}

\usepackage{graphicx}
\usepackage{chemformula}
\usepackage[margin=25mm]{geometry}
\usepackage{hyperref}
\usepackage{cleveref}
\usepackage{booktabs}
\usepackage{subcaption}
\usepackage[print-unity-mantissa=false]{siunitx}
\usepackage{newunicodechar}
\newunicodechar{κ}{$\kappa$}

\newcommand{\fom}[1][]{\Phi_\text{M#1}}
\newcommand{\CsTaTeO}[1][]{\ch{CsTaTeO6#1}}
\newcommand{\BiZrO}[1][]{\ch{Bi2Zr2O7#1}}
\newcommand{\A}{\protect\text{Å}} %
\newcommand{\egap}{{E_\text{gap}}}
\newcommand{\eform}{E_\text{form}}
\newcommand{\epstot}{\epsilon_\text{tot}}
\newcommand{\epsreal}{\epsilon_\text{real}}
\newcommand{\epsimag}{\epsilon_\text{imag}}
\newcommand{\epsionic}{\epsilon_\text{ionic}}
\newcommand{\epselec}{\epsilon_\text{elec}}
\newcommand{\dielloss}{\tan(\delta)}

\title{
  Pushing the Pareto front of band gap and permittivity:\\
  ML-guided search for dielectric materials
}

\author{Janosh Riebesell, T. Wesley Surta, Rhys Goodall, Michael Gaultois, Alpha A Lee}
\date{\today}

\begin{document}
\maketitle

\begin{abstract}
    Materials with high-dielectric constant easily polarize under external electric fields, allowing them to perform essential functions in many modern electronic devices.
    Their practical utility is determined by two conflicting properties: high dielectric constants tend to occur in materials with narrow band gaps, limiting the operating voltage before dielectric breakdown.
    We present a high-throughput workflow that combines element substitution, ML pre-screening, ab initio simulation and human expert intuition to efficiently explore the vast space of unknown materials for potential dielectrics, leading to the synthesis and characterization of two novel dielectric materials, \CsTaTeO{} and \BiZrO{}.
    Our key idea is to deploy ML in a multi-objective optimization setting with concave Pareto front.
    While usually considered more challenging than single-objective optimization, we argue and show preliminary evidence that the $1/x$-correlation between band gap and permittivity in fact makes the task more amenable to ML methods by allowing separate models for band gap and permittivity to each operate in regions of good training support while still predicting materials of exceptional merit.
    To our knowledge, this is the first instance of successful ML-guided multi-objective materials optimization achieving experimental synthesis and characterization.
    \CsTaTeO{} is a structure generated via element substitution not present in our reference data sources, thus exemplifying successful de-novo materials design.
    Meanwhile, we report the first high-purity synthesis and dielectric characterization of \BiZrO{} with a band gap of 2.27 eV and a permittivity of 20.5, meeting all target metrics of our multi-objective search.

\end{abstract}

\section{Introduction}
\label{sec:introduction}

Dielectric materials are indispensable in numerous modern electronic devices including central processing units (CPUs), random access memory (RAM), solid-state disks (SSDs), high-frequency (5G) antennas, photovoltaics, and light-emitting diodes (LEDs) \cite{wang_high_2018,ponceortiz_highk_2010}.
Their utility hinges on the intricate balance between dielectric constant and band gap, two anti-correlated properties that rarely co-occur in a single material.
High band gaps are crucial for reducing leakage current and preventing dielectric breakdown when subjected to high voltage.
Conversely, a large dielectric constant is desirable for minimizing the energy required for polarization, which is especially important in applications like transistor gates.
As transistors continue to shrink, the need for materials that can serve as ultra-thin gate dielectrics while withstanding operating voltages grows.

Historically, the discovery of dielectric materials has often relied on trial and error.
Recent advancements, particularly in automated workflows for computational screening using density functional perturbation theory (DFPT) have shown promise in systematically searching for high-performance dielectrics, e.g. mapping the bandgap-dielectric Pareto front of binary and ternary oxides \cite{yim_novel_2015}.
Improvements in compute power and workflow robustness have since enabled the scaling to several thousand diverse materials \cite{petousis_high-throughput_2017, petretto_highthroughput_2018,choudhary_highthroughput_2020}.

However, the sheer size of the space of $\sim$\num{e5} known, let alone the $\sim$\num{e10} hypothesized materials (up to quaternary order) \cite{davies_computational_2016}, prohibits sampling without inductive bias and presents a daunting challenge for existing computational methods.
Consequently, the dielectric properties of the vast majority of the $\sim\num{e7}$ simulated inorganic crystals remain unknown, making it likely a more comprehensive exploration of the space should yield novel high-performance materials.
To screen even a small subset of the full space requires orders of magnitude cheaper methods.
Worse, to go beyond the \num{e5} known materials introduces another layer of computational complexity in the form of thermodynamic stability prediction on top of estimating band gap and dielectric constant.

To address this, we propose a new dielectric discovery workflow that judiciously integrates machine learning (ML) as the first filter in a multi-step funnel.
ML, while less reliable than traditional methods like DFPT, is orders of magnitude faster and quickly improving in accuracy.
Our ML-guided approach uses surrogate models for band gaps, dielectric constants, and formation energies.
Instead of exact Cartesian coordinates, we employ Wyckoff positions for a coordinate-free, coarse-grained crystal structure representation.
This enables rapid generation and stability prediction of novel structures through elemental substitutions.
Following DFPT validation of the most promising candidates, the last selection step is an expert committee to incorporate human intuition when weighing the risks, precursor availability and ease of experimental synthesis of all high-expected-reward materials.
Finally, we validate the whole workflow by deploying it from start to finish which culminated in making and characterizing two new metastable materials in the process: \CsTaTeO{} and \BiZrO{} which partially and fully satisfy our target metrics, respectively.

Finding exceptional materials that extremize a single property necessarily requires extrapolation from the training data, for example maximizing hardness \cite{zhang_finding_2021, zuo_accelerating_2021, schmidt_machinelearningassisted_2023}.
This is fundamentally at odds with the statistical nature of ML, leading to increased error and less reliable predictions.
Our approach diverges from previous efforts by choosing a target class of materials where the path to application relevance requires balancing multiple conflicting properties.
This allows ML models to operate within regions of good training support while still predicting materials with exceptional figures of merit.
This type of tradeoff is ubiquitous in material science and is seen in other materials classes such as thermoelectrics (need high low thermal but high electrical conductivity) \cite{gaultois_perspective_2016,yan_material_2015}, catalysts (need high activity for fast reactions which tends to lower selectivity, increasing unwanted side reactions) \cite{guan_bimetallic_2021,liutkova_ca_2023}, high-strength and shape-memory alloys (need high strength and high ductility) \cite{chiu_investigations_2022,li_effect_2021}, and many more.
While multi-objective optimization is often seen as compounding the discovery challenge, we propose that concave Pareto fronts such as the above examples may in fact facilitate ML-guided discovery by reducing the need for extrapolation.

Despite a nascent but growing body of work on automated and high-throughput synthesis \cite{king_rise_2011,bedard_reconfigurable_2018,steiner_organic_2019,burger_mobile_2020,szymanski_autonomous_2023,lunt_modular_2024}, experimental validation remains a key bottleneck in the design of materials.
The process of manually developing experimental synthesis recipes for theoretical materials is very time-consuming, often taking months to a year per material.
The central claim of ML-guided screening and related efforts in rational materials design is that we can reduce the downside risk of attempting novel synthesis procedures by increasing the hit rate of successful materials.
To test the performance of our ML-guided approach we developed synthesis procedures for two materials predicted to be high-performing - \BiZrO{} and \CsTaTeO{}, with the structure of \CsTaTeO{} coming from our generative workflow.
Both materials displayed dielectric character with measured permittivities in the 43rd and 81st percentile, respectively, of 136 experimental reference results for dielectric materials reported in \cite{petousis_high-throughput_2017,petousis_benchmarking_2016}, validating the benefits of our ML-guided workflow.

In summary, our work showcases an advancement in ML-guided materials discovery, demonstrating its potential in efficiently navigating the vast landscape of dielectric materials and balancing multiple material properties for optimal device performance.

\section{Results}
\label{sec:results}

We first report the computational output of our workflow and then present experimental validation of two novel dielectric materials, \CsTaTeO{} and \BiZrO{}.

\subsection{A scalable generative machine learning workflow for dielectric discovery}

\begin{figure}[htbp!]
    \centering
    \includegraphics[width=\columnwidth]{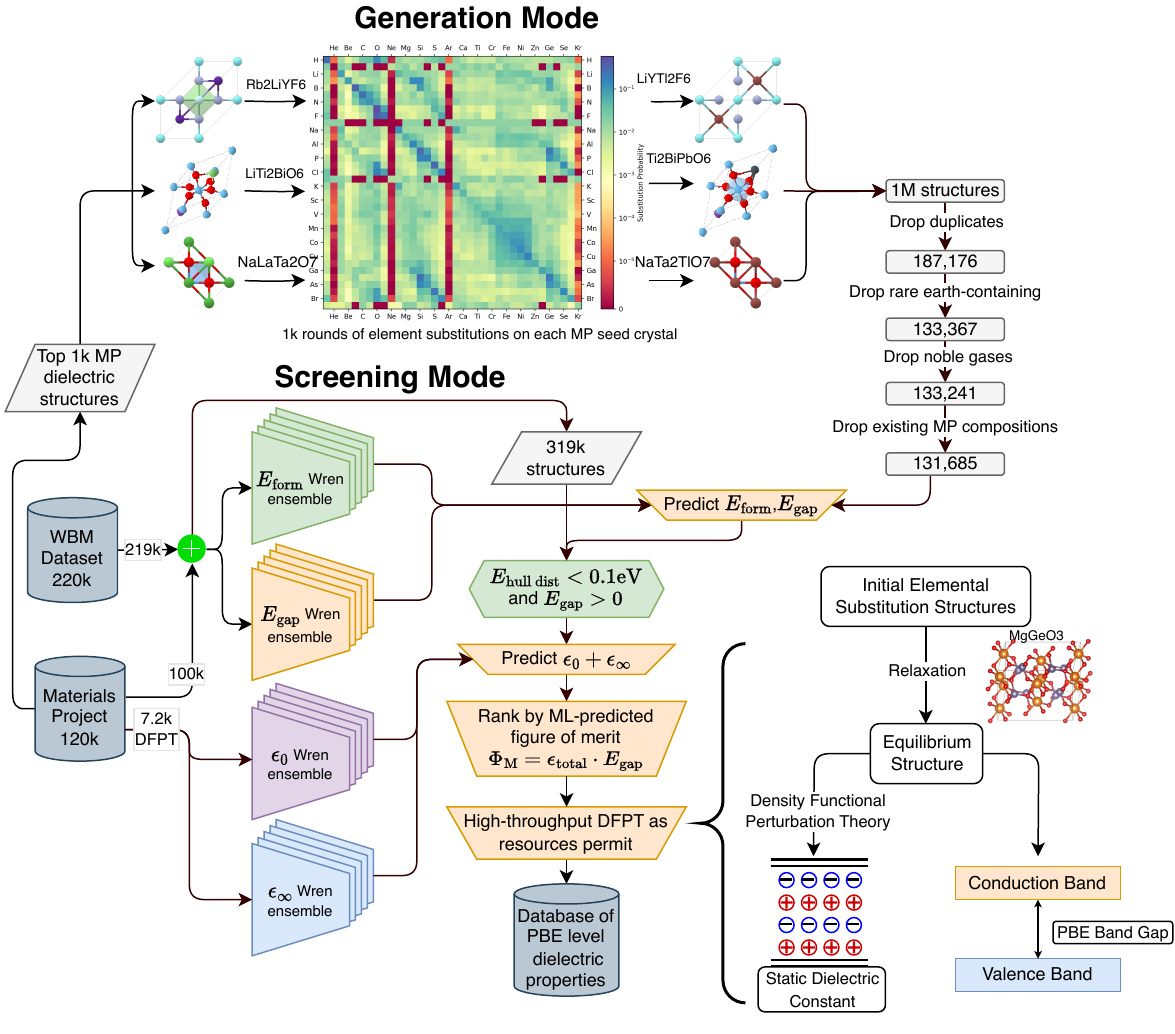}
    \caption{
        Diagram of our dielectric material discovery workflow, integrating ML pre-screening and elemental substitution for generating novel crystals with high-throughput DFPT validation. The discovery pipeline can operate in two modes: screening and generation. Screening mode searches for large permittivity among known materials. In generation mode, we feed the top 1k MP structures by figure of merit $\fom$ into an element substitution process.
    }
    \label{fig:discovery-workflow}
\end{figure}

This section describes the components and design decisions of our dielectric discovery workflow visualized in \cref{fig:discovery-workflow}.

The large search space and high cost of experimental validation demand a funnel approach to dielectric materials discovery.
To maximize the size of the initial candidate pool and still retain tractable computational cost, less auspicious materials must be discarded by a hierarchy of successively more expensive but higher-fidelity computational filters.
Such an approach maximizes return on invested effort by allotting more resources to candidates which accumulated evidence of expected utility in earlier filters.
Our proposed implementation for such a funnel workflow depicted in \cref{fig:discovery-workflow} precedes high-throughput DFPT with 5-6 orders of magnitude cheaper ML pre-screening to reduce a large list of \num{133241} candidate materials down to \num{2691} with computed dielectric properties.

We pre-screen based on 4 quantities - thermodynamic stability derived from predicted formation energy $\eform$, band gap $\egap$, ionic permittivity $\epsilon_0$ and electronic permittivity $\epsilon_\infty$ - each of which is predicted by a separate ensemble of 10 Wren models \cite{goodall_rapid_2021} independently trained from random initializations.
This allows us to both massively expand the search pool of initial candidates and waste fewer resources on unpromising compounds.
The formation energy and band gap training sets each consist of \num{319 601} data points, the combination of \num{98 850} Materials Project (MP) \cite{jain_commentary_2013} calculations and \num{220 751} from the WBM dataset \cite{wang_predicting_2021} (named WBM from the author's last name initials) which was generated with MP-compatible VASP settings.
The $\epsilon_0$ and $\epsilon_\infty$ ensembles are trained on the much smaller dataset of \href{https://materialsproject.org/materials?has_props=dielectric}{\num{7172} DFPT calculations in MP} (database version 2020-09-08) due to the lack of additional MP-compatible dielectric datasets.

While simply screening materials within large ab-initio databases for which properties of interest have yet to be calculated is a viable strategy, it is also important to demonstrate the generative capabilities of ML-based workflows.
To this end, we identify the top 1k MP structures by figure of merit $\fom = \epstot \cdot E_\text{gap\ PBE}$ and use them as seed crystals for element substitution.
The expectation is that this generates novel structures with increased likelihood of high $\fom$.
The substitution process involves replacing all sites of one element in the structure with a chemically similar element (e.g. $\ch{Na} \to \ch{K}$), as determined by a similarity matrix mined from the ICSD \cite{bergerhoff_inorganic_1983}.
After filtering out duplicates (compositions that already exist in MP or WBM, i.e. we do not consider structural degrees of freedom) as well as compounds containing noble gases, lanthanides or actinides, we are left with \num{131685} potential new dielectric materials.

Using the trained Wren ensembles, we predict $\eform$, $\egap$, $\epsionic$ and $\epselec$ for all candidates, both those sourced from high-throughput databases and those produced using our generative methodology.
We estimate the convex hull distance for each crystal from these predicted energies and discard those more than \SI{0.1}{eV/atom} above the hull.
This is motivated by the observation that 90\% of crystals in ICSD are predicted to be less than \SI{0.067}{eV/atom} above the convex hull \cite{sun_thermodynamic_2016}.
This tolerance towards instability accounts for errors in DFT energies and the fact that some thermodynamically unstable materials are kinetically or entropically meta-stable and hence synthesizable.

The remaining candidates are ranked by their ML-predicted figure of merit $\fom^\text{Wren}$ and subjected to a high-throughput DFPT workflow as our computational budget permits, resulting in a database of \num{2691} dielectric properties.

\subsection{Computational discovery of dielectrics beyond the Pareto front}
\label{sec:computational-results}

\begin{figure}
    \centering
    \includegraphics[width=0.97\linewidth]{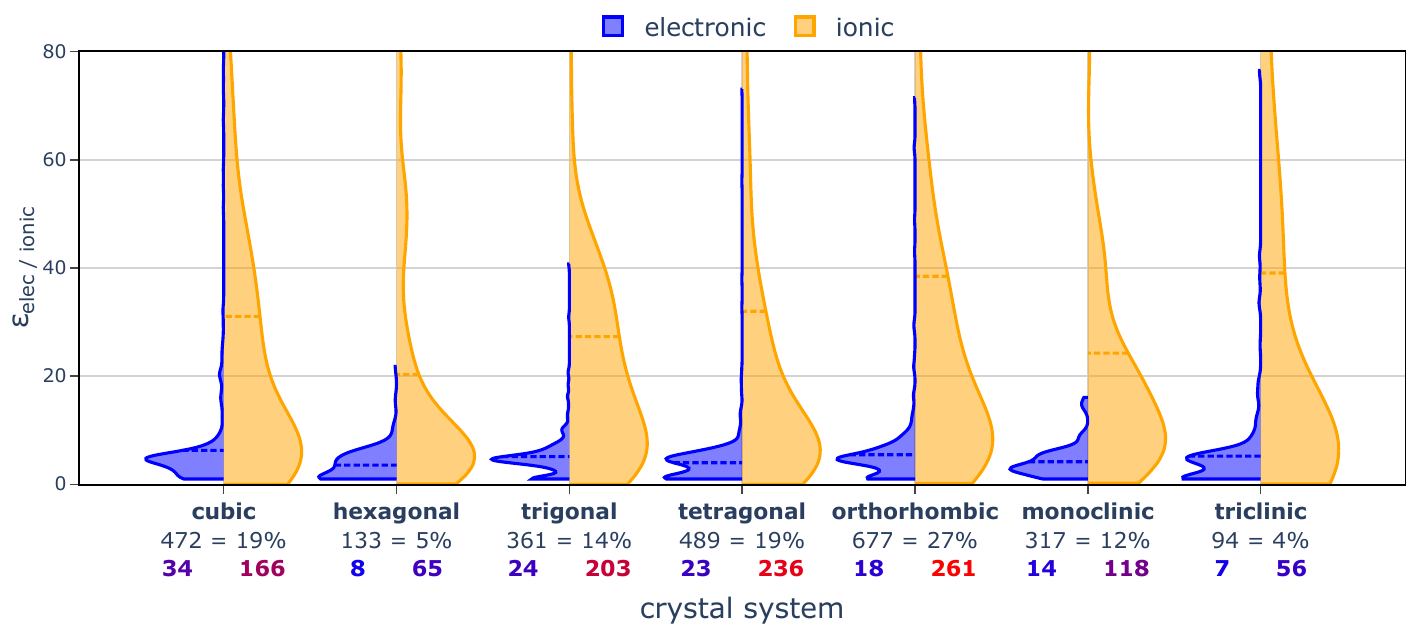}
    \caption{
        Violin plot showing Gaussian KDEs of DFPT-computed electronic (blue left halves) and ionic (orange right halves) contributions to the dielectric constant split by crystal system.
        The dashed horizontal lines in each violin show the median.
        Below each crystal system is the number of materials we have for it as well as its share of the total DFPT dataset in percent.
        The colored bold numbers (blue = low, red = high) show the mean of the top 30 electronic/ionic dielectric constants for each crystal system.
    }
    \label{fig:our-diel-elec-vs-ionic-violin}
\end{figure}

The violin plot in \cref{fig:our-diel-elec-vs-ionic-violin} shows Gaussian kernel density estimates (KDE) of all \num{2691} DFPT-computed electronic and ionic dielectric constants split by crystal system.
Unlike the electronic contribution which is lower-bounded by the vacuum permittivity, the ionic dielectric constant can be zero in all crystal systems.
We observe a general trend of higher dielectric constant the higher the crystal symmetry, especially for the ionic contribution.
Only cubic crystals reach significant electronic permittivity with a median of 10.

\begin{figure}[t]
    \centering
    \includegraphics[width=0.8\linewidth]{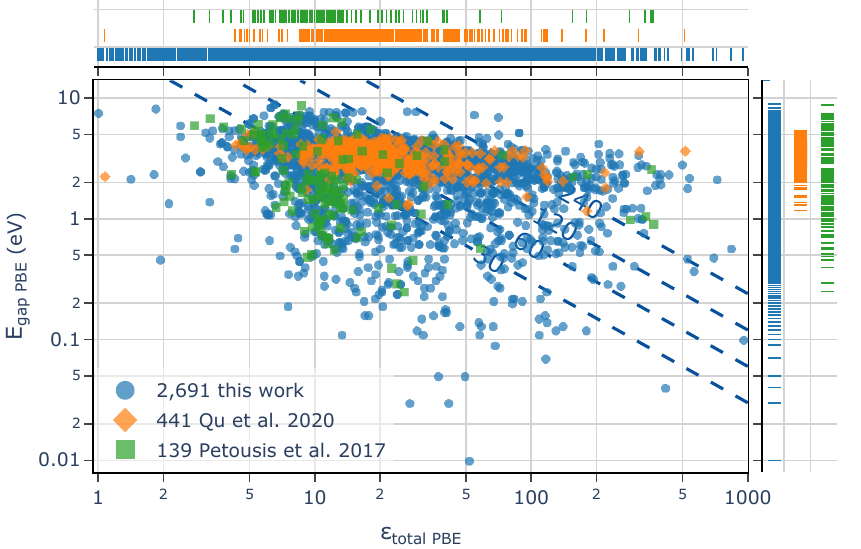}
    \caption{
        Log-log plot of PBE band gap $\egap$ vs. total dielectric constant $\epstot$ visualizing the hit rates for high-$\fom$ materials from different studies.
        Many of our DFPT data points (blue circles) reach into regions far beyond the \SI{240}{eV} isoline.
        The orange diamonds and green squares show results from \citeauthor*{petousis_high-throughput_2017}\cite{petousis_high-throughput_2017} and \citeauthor*{qu_high_2020}\cite{qu_high_2020} which produce fewer $\fom > 240$ materials, both in absolute numbers and as a fraction of dataset size (see \cref{tab:hit-rate-comparison}).
        The dark blue lines indicate constant figure of merit $\fom = \egap \cdot \epstot$.
        The stacked marginal rugs along the top and right show the distribution of band gaps and dielectric constants in each dataset.
    }
    \label{fig:pareto-us-vs-qu-vs-petousis}
\end{figure}

\Cref{fig:pareto-us-vs-qu-vs-petousis} compares the results from our methodology against those published in Petousis et al. \cite{petousis_high-throughput_2017} and Qu et al. \cite{qu_high_2020} by plotting the PBE band gap on the $y$-axis against the total dielectric constant on the $x$-axis on a log-log scale.
The blue circles show the \num{2691} DFPT results we computed.
The 441 orange diamonds show data generated by \cite{qu_high_2020} while the 139 green squares are from \cite{petousis_high-throughput_2017}.
The dark blue dashed isolines indicate constant figure of merit at values $\fom = \egap \cdot \epstot = c \in \{30, 60, 120, 240\}$ for band gap $\egap$ and total dielectric constant $\epstot$.
Our results achieve a larger number of materials beyond the highest $\fom$ isoline of 240 than both previous works combined.
We also achieve a higher hit rate per DFPT calculation of such high-merit materials as shown in \cref{tab:hit-rate-comparison}.
For \citeauthor*{qu_high_2020} $15 / 441 = 3.4\%$ of materials achieve $\fom > 240$ , while \citeauthor*{petousis_high-throughput_2017} reach $7 / 139 = 5.0\%$ and our data has $155 / 2680 = 5.8\%$ materials with $\fom > 240$.
Note that our hit rate increases even further when post-hoc excluding metals, i.e. filtering the hit rate analysis for materials with a band gap of at least 0.1 eV.
While the other works started from DFT structures with known band gaps and hence were able to filter out metals from the outset, the same is not possible when generating novel crystals with unknown electronic structures.
Our workflow instead relies on ML band gap prediction to filter out metals.
This step unfortunately suffers from a high false positive rate (metals misclassified as semiconductors/insulators).
Thus by upgrading to a better band gap model, a future realization of our workflow could achieve a high-merit hit rate in excess of $154 / 2063 = 7.5\%$.

\begin{table}[ht!]
    \centering
    \caption{Hit rate comparison for materials with $\fom > 240$. Excluding metals misclassified as insulators by our band gap models (which did not enter the other works in the first place), we achieve a $\fom > 240$ hit rate of 7.5\%. This validates our approach of creating candidate structures from known dielectrics and pre-screening with ML.}
    \begin{tabular}{lcc}
        \hline
        Study                                & Number of Hits / Total & Hit Rate (\%)  \\
        \hline
        Petousis et al.
        \cite{petousis_high-throughput_2017} & $7 / 139$              & $5.0$          \\
        Qu et al.
        \cite{qu_high_2020}                  & $15 / 441$             & $3.4$          \\
        This work                            & $155 / 2,691$          & 5.8            \\
        This work (with $ \egap > 0.1 $ eV)  & $154 / 2,067$          & $\textbf{7.5}$ \\
        \hline
    \end{tabular}
    \label{tab:hit-rate-comparison}
\end{table}

\Cref{fig:pareto-us-vs-qu-vs-petousis} compares our DFPT data to the results of \citeauthor*{petousis_high-throughput_2017}\cite{petousis_high-throughput_2017} and \citeauthor*{qu_high_2020}\cite{qu_high_2020}.
Our workflow generates more high-$\fom$ materials than both previous works combined and at a higher hit rate per expensive DFPT calculation than either \citeauthor*{petousis_high-throughput_2017}\cite{petousis_high-throughput_2017} or \citeauthor*{qu_high_2020}\cite{qu_high_2020}.
We believe this hit rate increase is attributable to ML pre-screening and substituting elements into known dielectric materials.

\subsection{Prospective Experimental Validation}
\label{sec:experimental-results}

To validate our workflow's ability to procure viable dielectric materials in practice, we selected \CsTaTeO{} and \BiZrO{} for experimental synthesis and characterization.
Our selection criteria incorporated DFPT results, prior literature or related materials appearing in the ICSD \cite{zagorac_recent_2019}, as well as precursor availability and expected ease of synthesis.
The selection process was facilitated by a custom web interface to visualize DFPT results on the Pareto front hooked up to a shared database for note-taking and collecting prior literature appearances on individual candidate materials detailed in \cref{sec:web-interface}.
Even so, making \CsTaTeO{} and \BiZrO{} required several trial-and-error iterations to optimize the synthesis conditions which we detail in this section.

\subsubsection{Optimization and Purity}

\begin{figure}[ht]
    \centering
    \begin{subfigure}[b]{0.4\linewidth}
        \includegraphics[width=\linewidth]{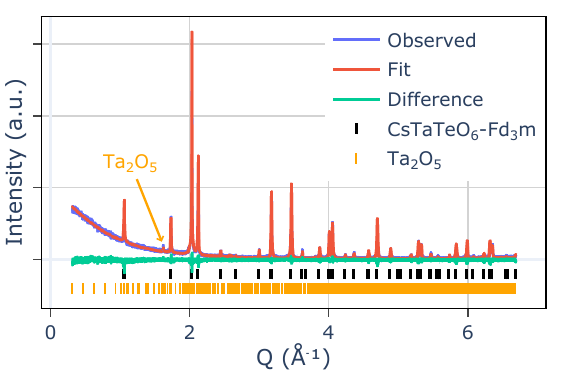}
        \caption{Pyrochlore Fd3m Rietveld fit for \CsTaTeO{}}
        \label{fig:exp-rietveld-CsTaTeO6-Fd3m}
    \end{subfigure}
    \hfil
    \begin{subfigure}[b]{0.42\linewidth}
        \includegraphics[width=\linewidth]{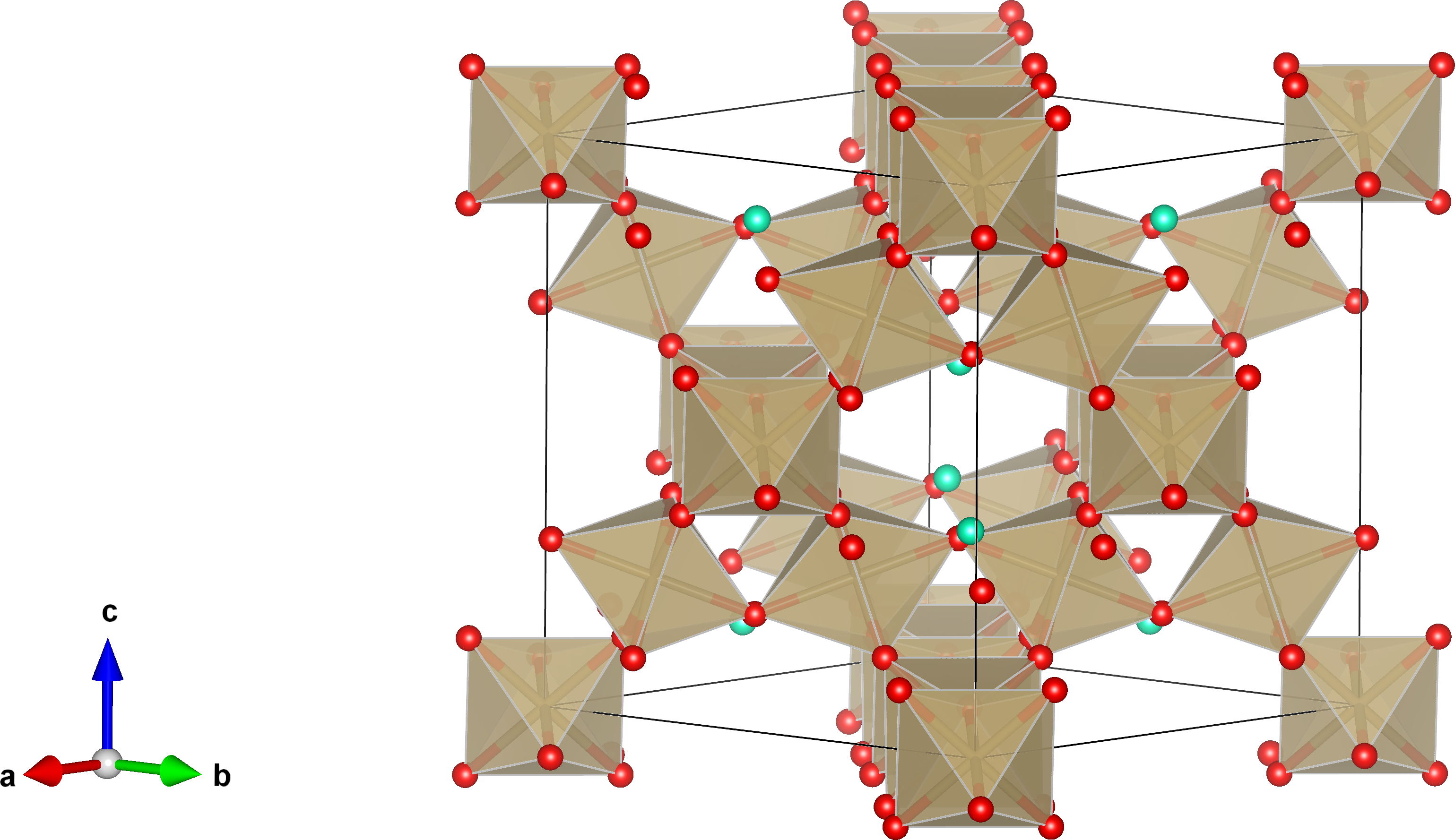}
        \caption{\CsTaTeO{} structural model}
        \label{fig:CsTaTeO6-whole-cell}
    \end{subfigure}
    \hfil
    \begin{minipage}[b]{0.16\linewidth}
        \begin{subfigure}[b]{\linewidth}
            \includegraphics[width=\linewidth]{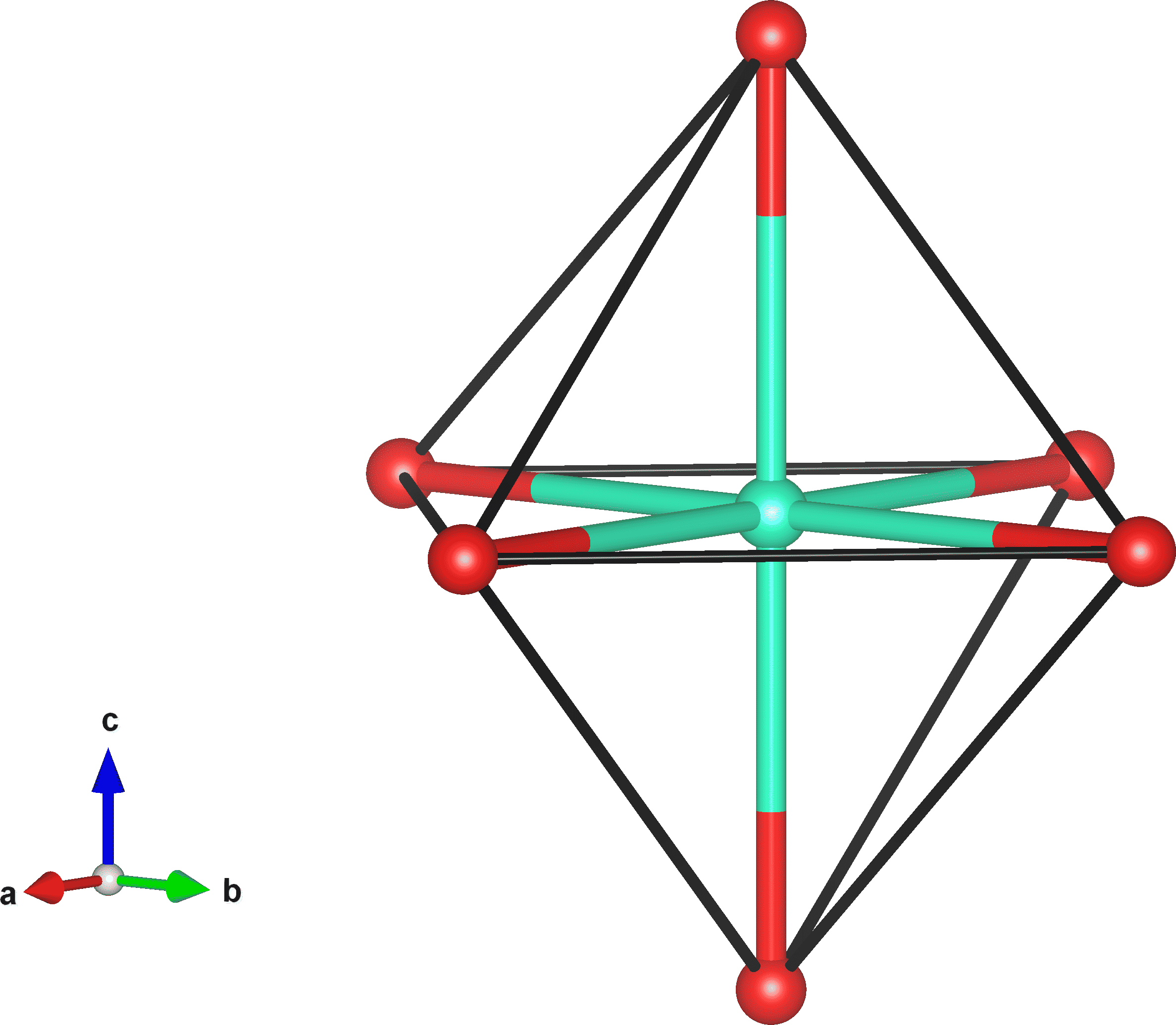}
            \caption{Pyrochl. A site}
            \label{fig:CsTaTeO6-a-site}
        \end{subfigure}
        \vfill
        \begin{subfigure}[b]{\linewidth}
            \includegraphics[width=\linewidth]{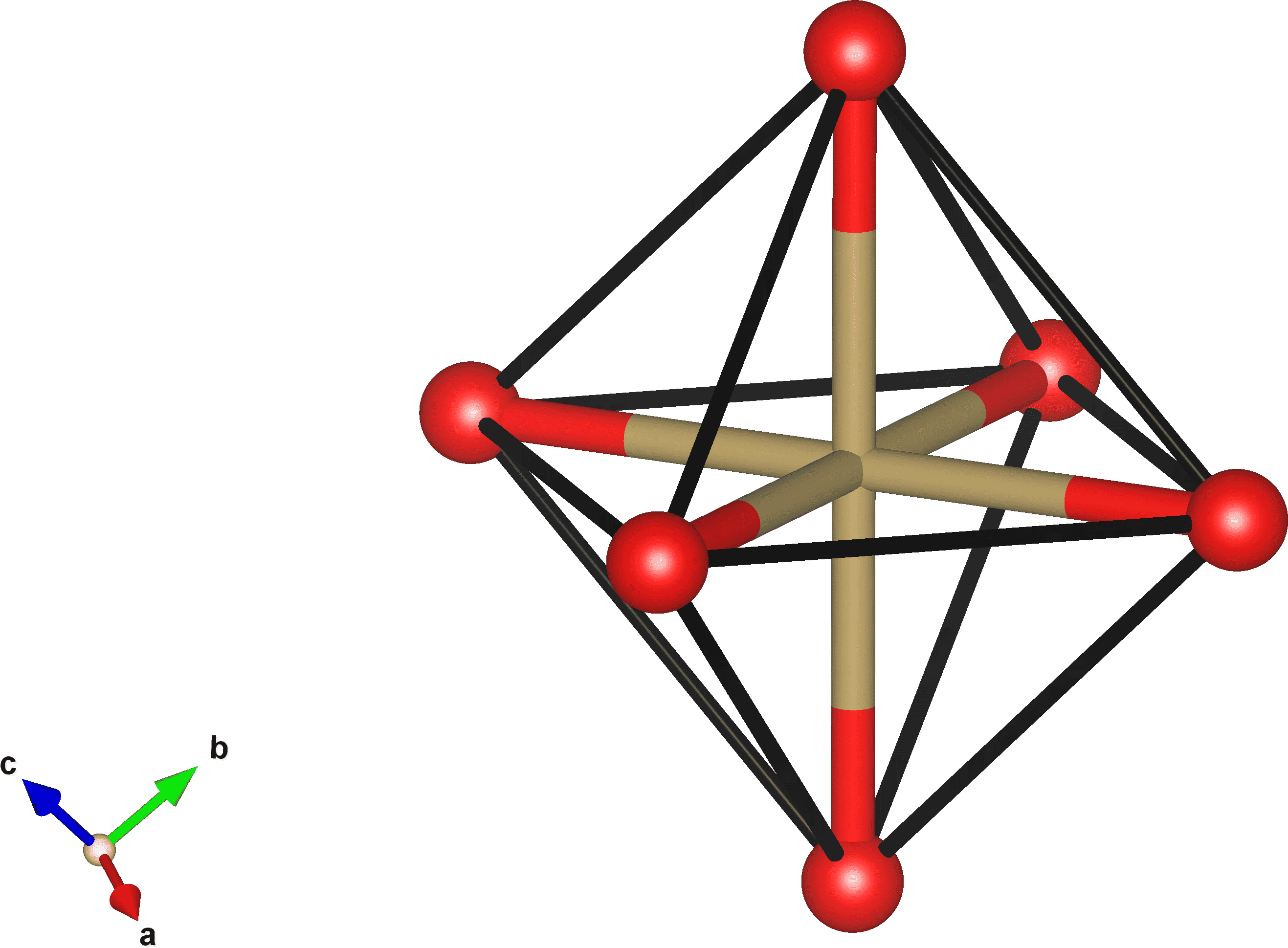}
            \caption{Pyrochl. B site}
            \label{fig:CsTaTeO6-b-site}
        \end{subfigure}
    \end{minipage}
    \begin{subfigure}[b]{0.4\linewidth}
        \includegraphics[width=\linewidth]{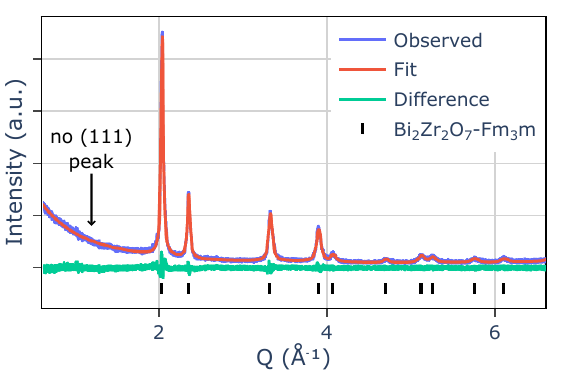}
        \caption{Fluorite Fm3m Rietveld fit for \BiZrO{}}
        \label{fig:exp-rietveld-Bi2Zr2O7-Fm3m}
    \end{subfigure}
    \hfil
    \begin{subfigure}[b]{0.28\linewidth}
        \includegraphics[width=\linewidth]{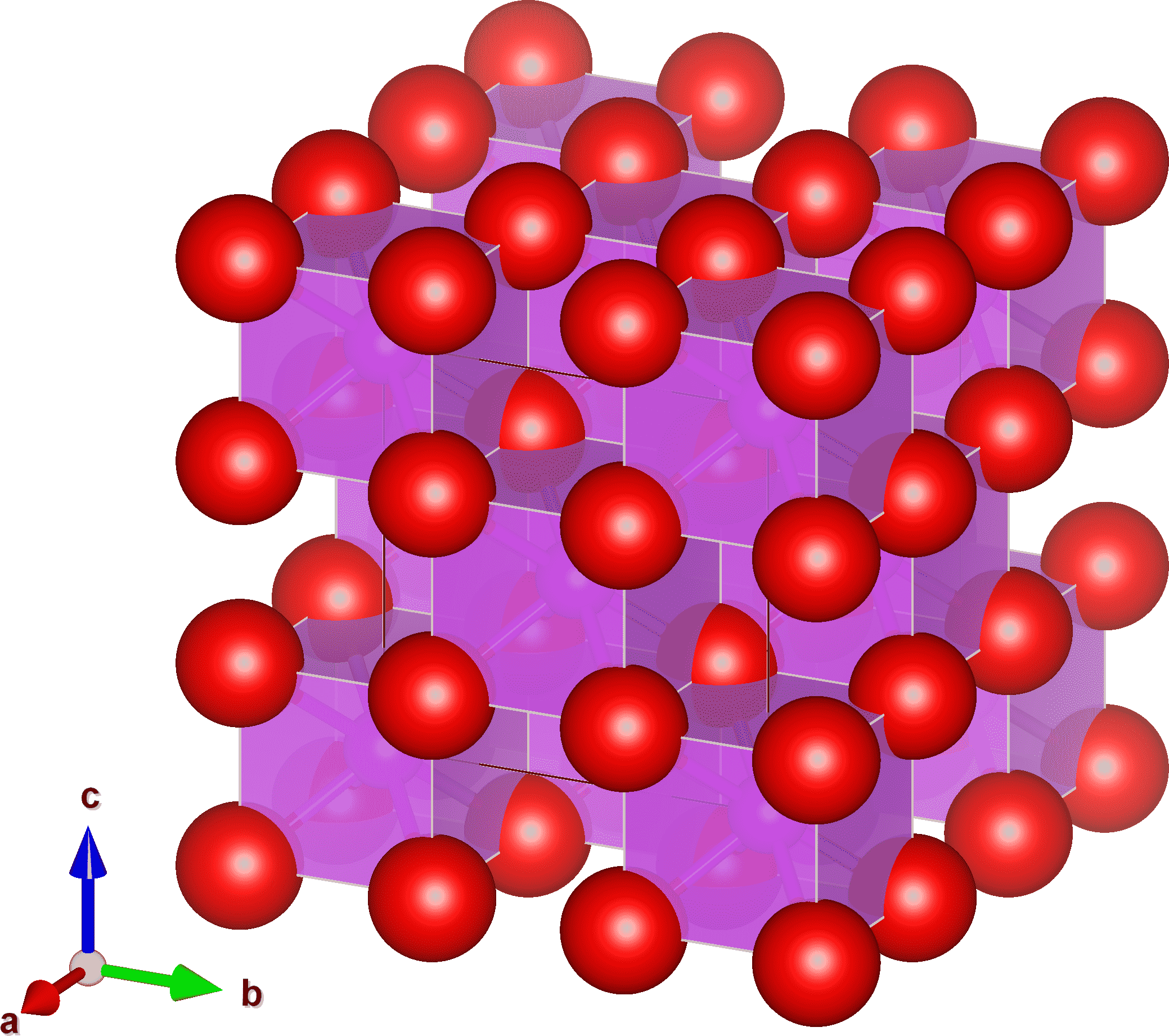}
        \caption{\BiZrO{} structural model}
        \label{fig:Bi2Zr2O7-whole-cell}
    \end{subfigure}
    \hfil
    \begin{subfigure}[b]{0.3\linewidth}
        \includegraphics[width=\linewidth]{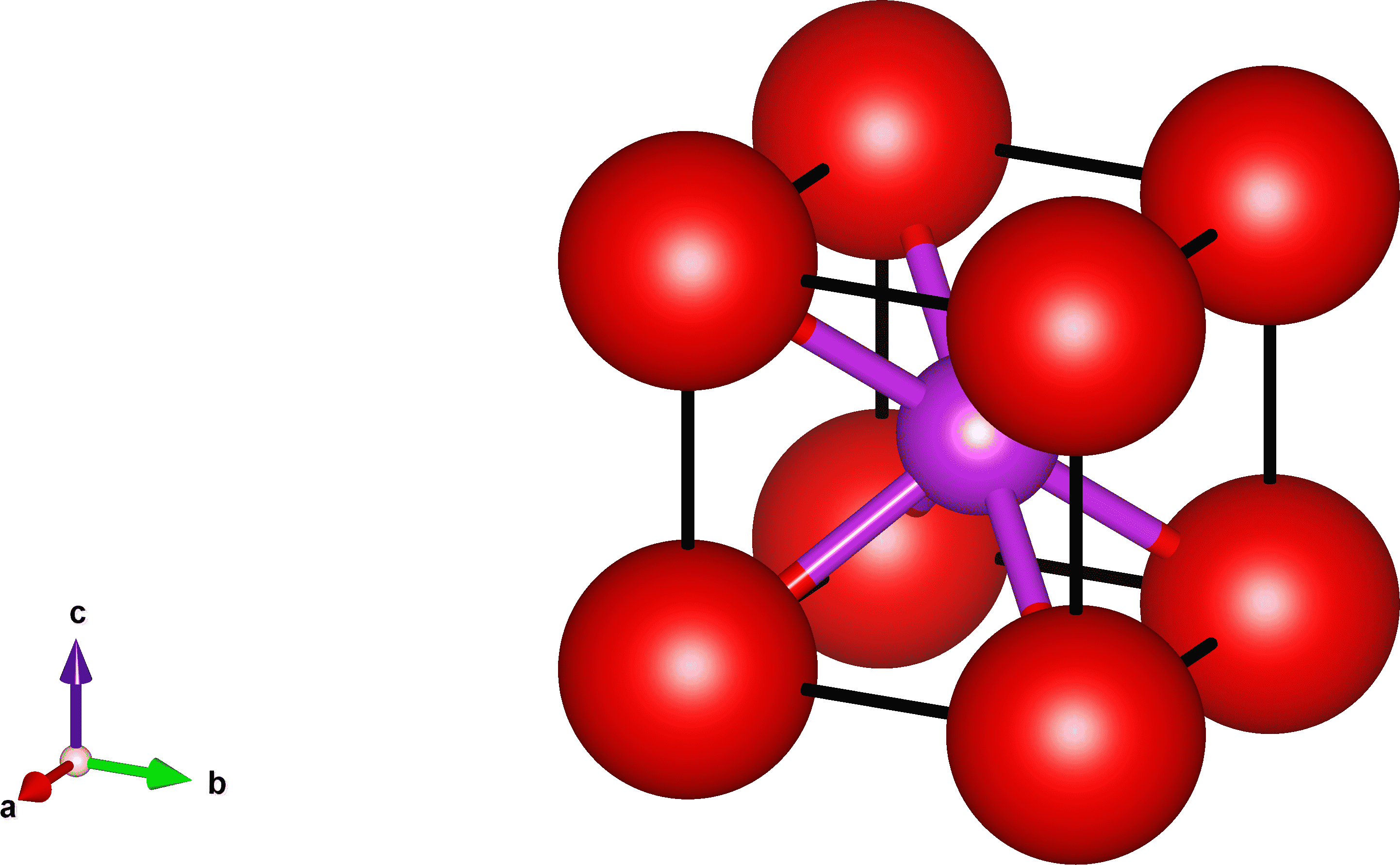}
        \caption{Isolated \ch{Zr/BiO8} polyhedra}
        \label{fig:Bi2Zr2O7-polyhedra}
    \end{subfigure}
    \caption{
        Structural determination of \CsTaTeO{} (\subref*{fig:exp-rietveld-CsTaTeO6-Fd3m}) and \BiZrO{} (\href{https://materialsproject.org/materials/mp-756175}{mp-756175}) (\subref*{fig:exp-rietveld-Bi2Zr2O7-Fm3m}) using XRD and Rietveld refinement.
        $Q = 2 \pi \cdot d^{-1}$ $[\A^{-1}]$ is the scattering vector.
        \subref*{fig:CsTaTeO6-whole-cell}) Crystal structure of the best Rietveld fit for \CsTaTeO{}.
        with \subref*{fig:CsTaTeO6-a-site}) and \subref*{fig:CsTaTeO6-b-site}) showing the pyrochlore A and B site octahedra.
        \subref*{fig:Bi2Zr2O7-whole-cell}) Crystal structure of the best Rietveld fit for \BiZrO{} with \subref*{fig:Bi2Zr2O7-polyhedra}) showing the isolated \ch{Zr/BiO8} polyhedra.
        Notable \ch{Ta2O5} impurities were detected in the \CsTaTeO{} XRD scan (\subref*{fig:exp-rietveld-CsTaTeO6-Fd3m}).
        \ch{Ta2O5} has many $hkl$ reflections, most of which are not distinguishable from the background noise.
        The most prominent observable \ch{Ta2O5} peak at $Q = 1.7$ as marked by the orange arrow.
        The absence of a (111) peak in the \BiZrO{} Rietveld fit (\subref*{fig:exp-rietveld-Bi2Zr2O7-Fm3m}) suggests a fluorite structure, in contrast to the literature-proposed pyrochlore model.
    }
    \label{fig:exp-structure-determination}
\end{figure}

\begin{figure}[ht]
    \centering
    \begin{subfigure}[b]{0.49\linewidth}
        \includegraphics[width=\linewidth]{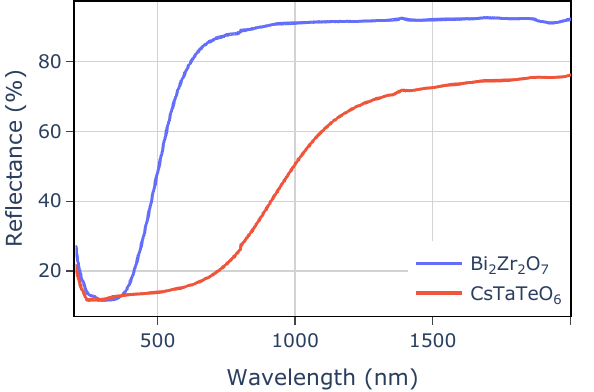}
        \caption{Diffuse reflectance spectra}
        \label{fig:exp-diffuse-reflectance}
    \end{subfigure}
    \hfil
    \begin{subfigure}[b]{0.49\linewidth}
        \includegraphics[width=\linewidth]{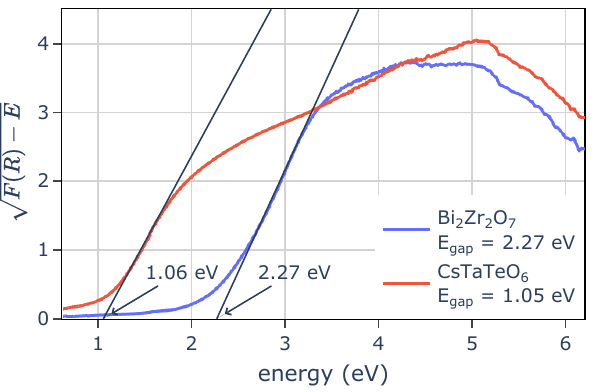}
        \caption{Optical band gap from Tauc}
        \label{fig:exp-tauc-bandgaps}
    \end{subfigure}
    \hfil
    \begin{subfigure}[b]{0.495\linewidth}
        \includegraphics[width=\linewidth]{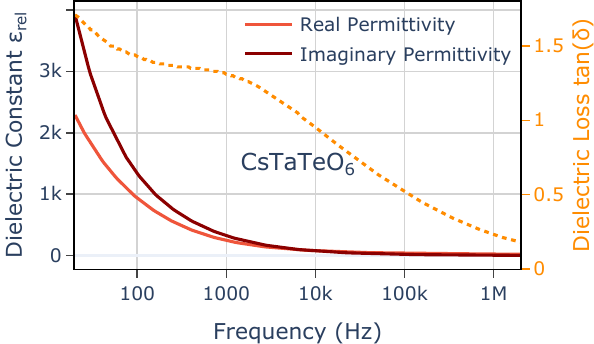}
        \put(-100, 68){\includegraphics[width=0.3\linewidth]{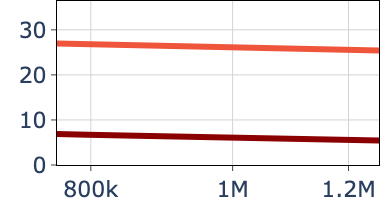}}
        \caption{\CsTaTeO{} dielectric response vs frequency}
        \label{fig:exp-CSTaTeO6-diel-real-imag-loss-vs-freq}
    \end{subfigure}
    \hfil
    \begin{subfigure}[b]{0.495\linewidth}
        \includegraphics[width=\linewidth]{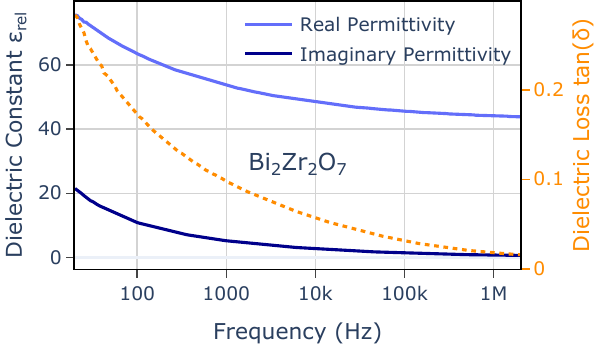}
        \caption{\BiZrO{} dielectric response vs frequency}
        \label{fig:exp-Bi2Zr2O7-diel-real-imag-loss-vs-freq}
    \end{subfigure}
    \caption{
        Dielectric measurements of \BiZrO{} and \CsTaTeO{}.
        (\subref*{fig:exp-diffuse-reflectance}) Diffuse reflectance spectra for both compounds exhibit distinctive absorption edges, indicating ordered crystalline structures.
        (\subref*{fig:exp-tauc-bandgaps}) Tauc plot measuring absorption coefficient $\alpha(E_\text{ph})$ vs photon energy $E_\text{ph} = h \nu$ for both compounds.The extracted optical band gaps are $\egap = \SI{2.27}{eV}$ for \BiZrO{} and \SI{1.05}{eV} for \CsTaTeO{}.
        (\subref*{fig:exp-CSTaTeO6-diel-real-imag-loss-vs-freq}) Dielectric response of \CsTaTeO{} as a function of frequency. We measure $\epstot = 26$ at \SI{1}{MHz} electric field (compared to 67 from DFPT)
        Its unwelcome high dielectric loss of $\dielloss = 0.23$ at \SI{1}{MHz} confirms the semiconducting nature observed in the Tauc plot's spectroscopic data.
        (\subref*{fig:exp-Bi2Zr2O7-diel-real-imag-loss-vs-freq}) Dielectric response of \BiZrO{} as a function of frequency yields $\epstot = 20.5$ at \SI{1}{MHz} (compared to 206 from DFPT)
        We highlight \BiZrO{}'s dielectric loss of less than 0.1 above \SI{1}{kHz}, a sufficiently low value for many practical applications.
    }
    \label{fig:exp-diel-characterization}
\end{figure}

We use X-ray diffraction (XRD) data and Rietveld fits to test our structural models for \CsTaTeO{} and \BiZrO{}.

The measured XRD pattern for our \CsTaTeO{} sample at 80\% of the theoretical weight density readily fits a pyrochlore model.
The atomic displacement parameters were small but positive within error.
Even though multiple disordered models were explored, the simplest pyrochlore provided the best fit.
We detected \ch{Ta2O5} impurities constituting $4.20 \pm 0.12$ \% of total weight that are highlighted in \cref{fig:exp-rietveld-CsTaTeO6-Fd3m}.

For \BiZrO{}, we explored optimal synthesis temperatures between \SIrange{550}{750}{\celsius}.
An extensive 8-hour XRD scan of \BiZrO{} after \SI{48}{h} of heating at \SI{650}{\celsius} confirmed the absence of \ch{Bi2O3} and \ch{ZrO2} impurities in the sample, which significantly surpasses existing literature in terms of purity \cite{pandey_metastable_2018}.
After sintering, we obtained a ceramic sample with 92+\% of the theoretical density of a single crystal.
Contrary to literature reports that typically describe an impure pyrochlore with a noticeable (111) reflection \cite{pandey_metastable_2018}, our samples exhibit no such peaks.
Prolonged heating did result in a broad (111) peak but was accompanied by undesired \ch{Bi2O3} and \ch{ZrO2} impurities.
Avoiding prolonged heat, the Rietveld analysis in \cref{fig:exp-rietveld-Bi2Zr2O7-Fm3m} shows the (111) peak to be absent, favoring a fluorite model for \BiZrO{}, in contrast to the literature-proposed pyrochlore models.
The compound exhibited large atomic displacement parameters ($B_\text{iso}$) which may arise from two superimposed crystallographic positions or due to off-stoichiometry (occupancy).
Both commonly result in models with large atomic displacement parameters that simulate the distribution of electron density from these sites.
However, attempts to reduce atomic displacement using site splitting and occupancy refinement for disordered materials did not yield better fits.
Higher-quality diffraction data, e.g. from neutron scattering, would likely be required for more accurate modeling.

Further details on synthesis development, equipment used and XRD fitting for both \BiZrO{} and \CsTaTeO{} are provided in methods \cref{sec:synthesis-details} and \cref{sec:Bi2Zr2O7-synthesis-development,sec:CsTaTeO6-synthesis-development}.

\subsubsection{Dielectric Characterization}
After having targeted, synthesized in high purity, and confirmed the structures of \CsTaTeO{} and \BiZrO{}, we investigated their physical properties.
The band gaps of both materials were identified using UV-vis impedance spectroscopy on powders using diffuse reflectance and an integrating sphere.
These data can be seen in \cref{fig:exp-diffuse-reflectance} and they were modified and fit using the Kubelka Munk \cite{kubelka_article_1931} equation to extract the bandgap, seen in \cref{fig:exp-tauc-bandgaps}.
\Cref{fig:exp-diffuse-reflectance} shows diffuse reflectance measurements for \CsTaTeO{} and \BiZrO{} exhibiting distinctive absorption edges.
The extracted band gaps are \SI{2.27}{eV} for \BiZrO{} and \SI{1.05}{eV} for \CsTaTeO{}.
It is worth noting that the measurements for both \CsTaTeO{} and \BiZrO{} turned out much lower than the DFT-calculated values of \SI{2.09}{eV} and \SI{2.96}{eV} respectively.
This is surprising given PBE's tendency to underestimate experimental band gaps.
For \CsTaTeO{} this may be due to complex defect effects not captured by DFT arising from \ch{Cs} or \ch{Te} volatility \cite{weiss_photoinduced_2020}.
A more accurate ML band gap model that provides a more specific filter for metals and semiconductors would save future implementations of our workflow from spending unwarranted compute and lab time on semiconducting compounds like \CsTaTeO{}.
However, given the limitations of PBE observed for these materials it would be advisable to train the model on reference data obtained from higher levels of theory.

The low value of $\egap = \SI{1.05}{eV}$ for \CsTaTeO{} is consistent with its observed black color and unfortunately renders it unusable as a dielectric material.
The dielectric measurements in \cref{fig:exp-CSTaTeO6-diel-real-imag-loss-vs-freq} confirm a band gap-related high dielectric loss\footnote{The dielectric loss measures dissipation of electromagnetic energy propagating inside a dielectric material to heat. It is defined as the phasor in the complex plane between the real resistive (lossy) and imaginary reactive (lossless) components of the relative permittivity $\epsilon_\text{rel} = \epsreal + i \epsimag$ and is commonly given as the tangent of that angle, $\dielloss = \epsimag / \epsreal$.}.
It is worth noting that despite its low band gap, \CsTaTeO{} exhibits high polarizability of $\epsreal = 26$ at 1 MHz up to its low breakdown voltage.
However, its high dielectric loss of $\dielloss = 0.23$ at \SI{1}{MHz} confirms the semiconducting behavior observed in the spectroscopic data.
We also caveat the measured dielectric constant with the fact that 23\% loss makes the extracted $\epsreal$ value less reliable.

The \BiZrO{} compound has an observed band gap of \SI{2.27}{eV}, making it a useful dielectric.
Importantly, the observed band gap is \SI{0.27}{eV} (12.5\%) higher than the previously reported mixed phase \cite{pandey_metastable_2018} who report $\egap = \SI{2}{eV}$.
This suggests reduced defect states and further substantiates the high purity and distinct phase of our synthesized materials.
Dense ceramics were only accessible using spark plasma sintering, due to the metastable nature of the compound.
Room temperature dielectric properties as a function of frequency can be seen in \cref{fig:exp-Bi2Zr2O7-diel-real-imag-loss-vs-freq}.
Dielectric properties arise from a variety of different mechanisms: space charge, dipolar, ionic, and electronic polarization.
Measuring as a function of frequency allows mechanisms with slower response times, such as space charge polarization arising from ionic conductivity, to be isolated from more meaningful mechanisms.
At high frequency (\SI{1}{MHz}) the dielectric response shows a dielectric permittivity ($\epsreal$) of 44 and a dielectric loss of $\dielloss = 0.018$.
The low dielectric loss (\textless 0.1) indicates that the value of $\epsreal$ is free from conductive contributions.
The permittivity of 44 is similar to doped \ch{Bi2O3} with fluorite-related structures, such as a 10\% \ch{Ta^5+}-doped \ch{Bi2O3} with a $\epsreal$ of 42 \cite{valant_dielectric_2004}.
However, \BiZrO{} has a higher $\epsreal$ than \ch{HfO2} or \ch{ZrO2} ($\epsreal$ between 22-25) fluorites which are used as high-k dielectrics industrially \cite{choi_development_2011}, making it a worthwhile material to consider for real-world application.
Furthermore, the aqueous-based synthesis with low calcination temperature of \BiZrO{} presents promising opportunities for solution processing of dielectrics which are compatible with existing industrial MOSFET processing technologies.

\clearpage
\section{Methods}

\subsection{Derivation of \texorpdfstring{$\fom$}{FoM}}
\label{sec:choosing-fom}

Since dielectric constant and band gap are both crucial factors when considering electronic device applications, we measure materials by a figure of merit defined as
\begin{equation}
    \label{eq:fom}
    \fom
    = \egap \cdot \epstot
    \qquad\text{where}\qquad\epstot = \epsionic + \epselec.
\end{equation}
A product ensures materials exhibit at least intermediate levels of band gap \textit{and} permittivity.
This follows \citeauthor*{yeo_mosfet_2003}\cite{yeo_mosfet_2003} who define this semi-empirical expression for the leakage current through a MOSFET gate dielectric:
\begin{equation}
    J_G
    \propto\,\exp \left\{{-} {\frac{4\pi \sqrt {2q}}{h}} \cdot \left(m_\text{eff} \ \Phi_b \right)^{1 / 2} \epstot \cdot t_\text{ox,eq} \right\}
\end{equation}
with charge $q$, effective tunneling mass $m_\text{eff}$ of the electron or hole, injection barrier of the gate dielectric $\Phi_b$, and the \ch{SiO2}-equivalent-capacitance oxide thickness $t_\text{ox,eq} = (\epsilon_\text{\ch{SiO2}}/\epstot) \cdot t_\text{phys}$.
Increasing $(m_\text{eff} \ \phi_b)^{1/2} \ \epstot$ exponentially suppresses the tunneling current. Thus MOSFET device miniaturization requires materials that maximize this quantity.
The effective tunneling mass $m_\text{eff}$ and the carrier injection barrier $\phi_b$ are expensive to compute from first principles and out of reach for high throughput workflows.
\citeauthor*{hinkle_novel_2004}\cite{hinkle_novel_2004} therefore approximate their product as proportional to the band gap, $\egap \propto (m_\text{eff} \ \phi_b)^{1/2}$.
Increasing $\fom = \egap \cdot \epstot$ should therefore result in exponentially suppressed tunneling current.

\subsection{Initial Candidate Generation}
\label{sec:initial_candidate_generation}

As shown in \cref{fig:discovery-workflow}, we begin our discovery campaign by generating a large set of initial candidates.
The Materials Project currently holds \num{7172} materials with DFPT-calculated permittivity.
Starting with the 1000 highest FOM MP dielectric materials, we perform 1000 rounds of elemental substitution on each source structure.
Substitutions are guided by a chemical similarity matrix \cite{wang_predicting_2021} mined from the Inorganic Crystal Structure Database (ICSD) \cite{bergerhoff_inorganic_1983}, resulting in 1 million potential new structures.

The chemical similarity matrix offers a likelihood score for elemental substitution, based on their co-occurrence in the same space group in ICSD.
This approach is inspired by previous works \cite{glawe_optimal_2016,goodall_rapid_2021}.
During substitution, we swap out one element for another across the entire structure and limit ourselves to the 89 elements present in the Materials Project.
This process yields 187,176 potential candidates, which we then filter as follows:

\begin{enumerate}
    \item Remove duplicates: $\num{1e6} \to \num{187176}$
    \item Exclude structures containing rare earths (lanthanides and actinides): $\num{187176} \to \num{133367}$
    \item Exclude structures containing noble gases: $\num{133367} \to \num{133241}$
    \item Remove existing Materials Project compositions: $\num{133241} \to \num{131685}$
\end{enumerate}

We remove rare earths because DFT is well-known to struggle with the 4f electrons \cite{soderlind_groundstate_2014}, making any DFPT on such compounds less reliable. We filter noble gases because they are chemically inert and hence unlikely to occur in stable compounds. We remove structures with matching compositions in the Materials Project since many MP structures are sourced from the ICSD and hence the experimentally observed ground state.
As such, any structures we generate with polymorphs in MP have increased risk of being metastable at best.

\subsection{Training Data}
\label{sec:training-data}

We trained the Wren ensembles for formation energy and band gap on the combination of two large datasets:

\begin{itemize}
    \item The \textbf{Materials Project (MP) database} \cite{jain_commentary_2013} is a well-curated database of high-throughput DFT calculations.
          At time of access, MP contained 146,323 crystal structures (database version 2020-09-08 powered by \texttt{pymatgen} version 2022.0.8) \cite{persson_materials_2022}.
    \item Wang et al. \cite{wang_predicting_2021} calculated energies and properties for a large number of crystal structures generated from MP source structures via elemental substitution with chemically similar elements as pioneered in \cite{glawe_optimal_2016}.
          After substitution, the structures were relaxed using MP-compatible workflows.
          Using the author's initials, we refer to this as the \textbf{WBM data set}.
          After de-duplication and cleaning, WBM contains 220k structures.
\end{itemize}

Together, MP and WBM provide \num{319601} formation energies, \num{319601} band gaps. The Materials Project also contains dielectric properties for \num{7172} materials which we used to train Wren ensembles that predict ionic and electronic permittivity.

\subsection{Machine Learning}
\label{sec:ml}

To predict formation energy, band gap and permittivity in the ML pre-filtering step for each of the 131,685 generated candidate materials (\cref{sec:initial_candidate_generation}), we utilize Wren ensembles \cite{goodall_rapid_2021} which use a coarse-grained Wyckoff position-based material representation that discards exact atomic coordinates in favor of discrete, enumerable symmetry labels identifying groups of sites that map onto each other under the crystal's symmetry operations.
Each Wyckoff position is embedded into a vector space and concatenated with the crystal site's Matscholar element embeddings \cite{weston_named_2019} before being placed in a fully connected graph with all other Wyckoff sites.
Each node in the graph is then allowed to contextualize to its neighbors via multiple message-passing layers and finally mean-pooled to get a permutation- and relaxation-invariant, fixed-length, symmetry-aware crystal descriptor which is much cheaper to obtain than relaxed atom positions.
A simple feed-forward net with skip connections \cite{he_deep_2015} and ReLU \cite{nair_rectified_2010} activations then maps the Wren crystal embedding onto one or multiple target variables.
This featurization becomes more informative with higher symmetry in the structure.
For our use case of filtering out unrelaxed structures immediately after elemental substitution, its distinct advantage is invariance under structure relaxations as long as the relaxation does not affect the structure's symmetry (many DFT relaxations enforce keeping the initial symmetry throughout the relaxation, e.g.
by setting $\text{ISYM} > 0$ for VASP).

For each of the four material properties of interest -- formation energy, ionic and electronic dielectric constants, and band gap -- we train Deep Ensembles \cite{lakshminarayanan_simple_2016} of 10 independent Wren models.
Trained on identical data but with different initializations, these ensembles offer two advantages:
\begin{itemize}
    \item The ensemble average yields more reliable point estimates compared to single models.
    \item Ensemble variance allows us to assess epistemic model uncertainty, which we incorporate into a risk-aware figure of merit via error propagation.
          This reduces false positives at the cost of increased false negatives.
\end{itemize}
The ensemble-risk-aware figure of merit $\fom^\text{std-adj}$ including uncertainty propagation reads:
\begin{equation}
    \fom^\text{std-adj}
    = \sqrt{(\epstot \cdot \sigma_{\egap})^2
        + (E_\text{gap} \cdot \sigma_{\epstot})^2},
\end{equation}
where $\epstot^\text{Wren}$ and $\sigma_{\egap}$ are the Wren ensemble mean and standard deviation for the predicted total dielectric constant. Likewise $E_\text{gap}^\text{Wren}$ and $\sigma_{\epstot}$ are the ensemble mean/std. dev. for the predicted band gap.
We use the $\fom^\text{std-adj}$ (rather than the standard $\fom$) to rank element substitution structures for priority when allocating compute budget for DFPT calculations.

Moreover, for the formation energy ensemble, we also estimate aleatoric uncertainty, i.e. uncertainty that is inherent to the data, by using a ``robust'' loss function.
This loss requires changing the final output layer of each model to predict two numbers per sample.
The loss function interprets the first number as the predicted mean and the second as predicted aleatoric uncertainty.
This uncertainty enters the loss function as an attenuation term on the $L^p$ norm.
This allows the model to deweight the loss on predictions it attributes higher uncertainty to at the cost of incurring a higher regularization penalty.

\def\f{\mathbf{f}} \def\x{\mathbf{x}} \def\y{\mathbf{y}}
\begin{equation}
    \mathcal{L} = \frac{1}{N} \sum_{i=1}^N \frac{1}{2\sigma(\x_i)^2} ||\y_i - \f(\x_i)||^2 + \frac{1}{2} \log \sigma(\x_i)^2
\end{equation}
where $\x_i$ are the model inputs, $\y_i$ the corresponding target value, $\f(\x_i)$ the model predictions and $\sigma(\x_i)$ is the observation noise parameter, the second predicted by the model.
The observation noise is learned by the model as a function of the input, making the loss heteroscedastic (i.e. sample-dependent).
Thus the model can learn to deweight the standard $L^2$ loss in the first term by increasing the predicted observation noise.

We do not set a robust loss on the other 3 ensembles due to an increase in validation error which we did not observe for the formation energy ensemble.%

For all 4 Wren ensembles (formation energy, band gap, ionic + electronic dielectric constant), we adopt the same hyperparameters as \Citeauthor*{goodall_rapid_2021} \cite{goodall_rapid_2021} to which we refer for details on the model architecture. In summary, each ensemble member consists of 3 message passing layers, each with a single attention head.
Both parts of the soft-attention mechanism use single-hidden layers with 256 hidden units and LeakyReLU activation functions. The output network following the message-passing layers is a simple feed-forward net with skip connections and ReLU activation functions. Its 4 hidden layers have sizes 64, 256, 256, and 1, respectively.

\subsection{DFT Structure Relaxation}
\label{sec:dft-structure-relaxation}

We used the Vienna ab-initio Simulation Package (VASP) \cite{kresse_efficiency_1996,kresse_efficient_1996} in projector augmented wave (PAW) mode \cite{blochl_projector_1994} to relax artificial crystal structures generated via elemental substitution of known structure prototypes.
The exchange-correlation energy was computed in the generalized gradient approximation (GGA) \cite{langreth_beyond_1983} using the Perdew-Burke-Ernzerhof (PBE) functional \cite{perdew_rationale_1996}.
Input files were auto-generated by \texttt{pymatgen} \cite{ong_python_2013}.
To perform high-throughput DFT, we used the Materials Project workflow library \texttt{atomate} \cite{mathew_atomate_2017}, the job launcher, queue manager and progress monitor Fireworks \cite{jain_fireworks_2015}, and the automatic error handler Custodian \cite{ong_python_2013}.
Structures were relaxed until all interatomic forces fell below \SI{e-2}{eV/\A} and the total energy change between self-consistent field (SCF) cycles fell below \SI{e-7}{eV}.

\subsection{Dielectric Properties from DFPT}
\label{sec:dfpt-to-dielectric}

Candidates that pass our ML filters are fed into high-throughput density functional perturbation theory (DFPT).
This stage offers more accurate property estimates at 3-4 orders of magnitude increase in computational cost.

We computed the electronic permittivity using linear response theory at the generalized gradient approximation (GGA) level of density functional perturbation theory as implemented in VASP 6.2.1.
High-throughput calculations were orchestrated on the Cambridge CSD3 cluster using the \texttt{wf\_dielectric\_constant} workflow in \texttt{atomate} v1.1.0 \cite{mathew_atomate_2017}.
This yields Born effective charges and phonon modes at the $\Gamma$ point \cite{lee_comparative_2007}.

We depart from standard MP dielectric settings in several respects.
First, by using the (at the time) most recent \texttt{PBE\_54} release of VASP POTPAW pseudopotentials (MP uses \texttt{PBE}).
We used the default structure-dependent $k$-point grid as implemented in \texttt{pymatgen} which constructs Gamma-centered meshes for hexagonal and face-centered cells, and Monkhorst-Pack grids otherwise.
However, we increased the grid density to \num{3000} $k$-points per atom despite the significant cost increase for a high-throughput workflow to accommodate the sensitivity of linear-response calculations to $k$-point sampling.
We also set tight convergence criteria of \texttt{EDIFF} = \SI{e-7}{eV} (default = \SI{e-5}{eV}) and a high kinetic energy cutoff for the plane wave basis set of \texttt{ENCUT} = \SI{700}{eV} (default = \SI{520}{eV}).
We expect these changes to increase the fidelity of our results, or at worst, increase compute cost at no benefit.

The total dielectric tensor splits into ionic ($\epsilon^0$) and electronic ($\epsilon^\infty$) contributions:

\begin{equation}
    \epsilon_{ij}^\text{total}
    = \epsilon_{ij}^0+\epsilon_{ij}^\infty
\end{equation}
with $i,j \in {x,y,z}$ the 3 spatial dimensions and 0 and $\infty$ representing the electric field frequency.
The ionic contribution $\epsilon^0$ is computed from the Born effective charges $Z^*$ and the phonon modes $\omega$ \cite{gonze_dynamical_1997},

\begin{equation}
    \epsilon_{ij}^0
    = \frac{4 \pi}{\Omega} \sum_{m} \frac{Z_{m,i}^* Z_{m,j}^*}{\omega_m^2}
\end{equation}
with $\Omega$ the unit cell volume, $m$ the phonon mode index, $\omega_m$ the infrared phonon frequency of
mode $m$ and $Z_{m,i}^*$ the $i$th component of the Born effective charge of mode $m$.

The scalar dielectric constant that enters our figure of merit equation \cref{eq:fom} is the mean of the eigenvalues $\epsilon_i$ of the total dielectric tensor:

\begin{equation}
    \epstot
    = \frac{1}{3} \sum_{i=1}^3 \epsilon_i^\text{total}
\end{equation}
The ionic contribution $\epsionic$ to the permittivity is known to be sensitive to low-frequency phonon modes which are incorrectly softened by the lattice parameter overestimation typical for GGA \cite{yim_novel_2015}, resulting in higher mean error than LDA.
We chose to run GGA DFPT despite this known GGA shortcoming to retain compatibility with existing MP dielectric data \cite{petousis_benchmarking_2016, petousis_high-throughput_2017}.

\subsection{Web Interface for Collaborative Synthesis Selection}
\label{sec:web-interface}

\begin{figure}
    \centering
    \includegraphics[width=\linewidth]{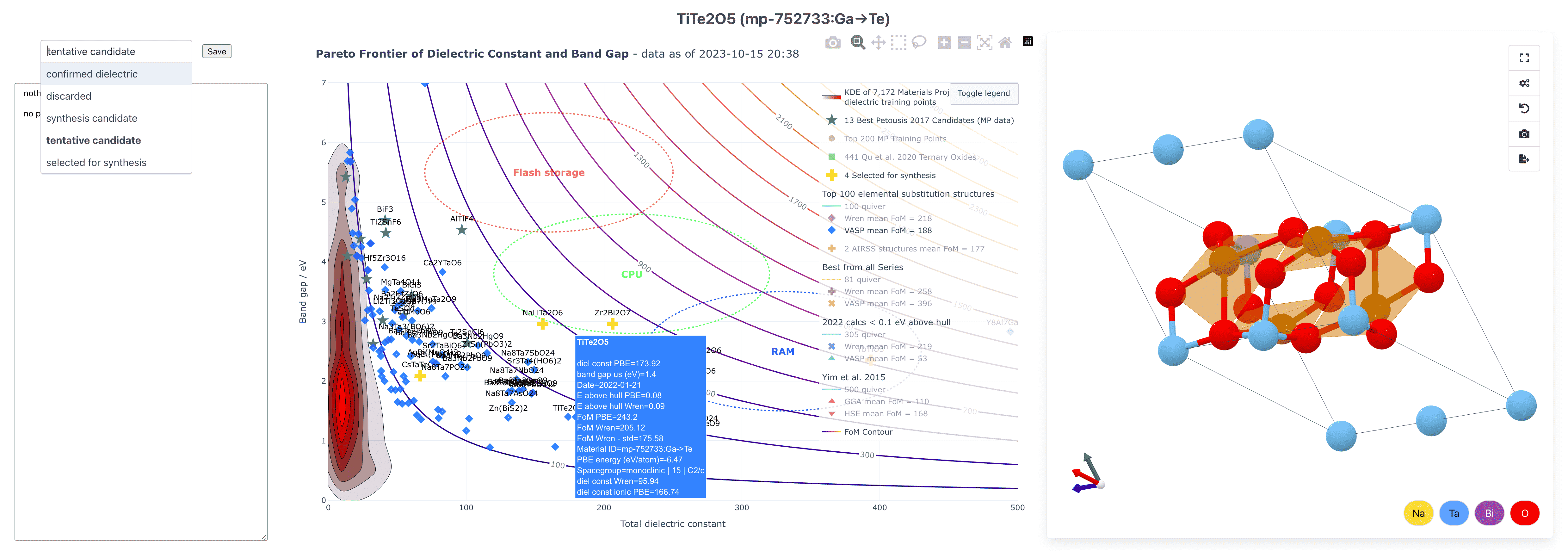}
    \caption{
        Screenshot of the \href{https://janosh.github.io/dielectrics}{web app} that aids with synthesis selection.
        The centerpiece of the app is an interactive scatter plot similar to \cref{fig:pareto-us-vs-qu-vs-petousis} showing the Pareto front of band gap and total dielectric constant. The legend on the right enables toggling between our own data set and data from prior works discussed in \cref{sec:related-work}. For our own data, we have legend groups for different calculation batches and $\fom$-based subsets of the data that allow switching between viewing ML predictions and/or their corresponding DFPT results with a third option to show a quiver plot that renders arrows between these points. This makes it easy to visualize how ML predictions differ from DFPT results and to look for trends in ML errors w.r.t. chemistry. Ellipses again indicate regions of particular interest for specific device applications (CPU, RAM, Flash storage). The density contours show lines of constant figure of merit.
    }
    \label{fig:plotly-web-app}
\end{figure}

The last step in our discovery workflow before experimental synthesis involves a custom-built web interface shown in \cref{fig:plotly-web-app} powered by a MongoDB Atlas M2 instance on the backend. This database is automatically updated when new \texttt{atomate} \cite{mathew_atomate_2017} DFPT workflows finish on our compute cluster. To implement the frontend, we leveraged multiple open-source technologies:

\begin{itemize}
    \item \texttt{pymongo} \cite{fedorova_writes_2022} for fetching the latest calculation results from our \texttt{atomate} \texttt{tasks} collection.
    \item \texttt{plotly} powers the interactive scatter plot used to show computed and/or ML-predicted band gaps and dielectric constants, switch between different calculation series or best-of subslices of the data as well as clicking points to select individual materials for closer inspection.
    \item \texttt{CrystalToolkit} \cite{horton_crystal_2023} renders the 3d structure of the material selected in the scatter plot with pan and zoom functionality.
    \item \texttt{dash} \cite{hossain_visualization_2019} stitches the above 3 components together with callback functions. The two main ones are updating the structure viewer whenever a new point is selected from the scatter plot and updating the database when new free-form notes or categorizations are recorded in the text area and dropdown menu on the left.
\end{itemize}

In our case, selecting individual points from the scatter plot and annotating them with free-form text took place during remote live discussions between theorists and experimentalists while screen sharing. Since these meetings took place over months, the web app massively helped with keeping track of reasons for categorizing a given material as discarded or tentative/firm synthesis candidate or recording links to prior art for materials categorized as already confirmed dielectrics.
Given this web app proved an enabler of effective remote collaboration between computational and experimental labs, we emphasize the importance of developing more custom tools that improve information flow and data visualization. Moreover, we found our tool significantly facilitates the process of keeping provenance. Ideally, this process should be automated entirely in the future as this area is extremely prone to human error.

The final verdict of these discussions results in a classification as one of
\begin{description}
    \item[confirmed dielectric]: prior experimental literature exists confirming our candidate material to be a dielectric. No point in synthesizing and re-characterizing, but increases trust in our workflow.
    \item[selected for synthesis]: Promising in every way, i.e. high calculated band gap and permittivity, cheap and easily accessible precursors, synthesis procedure matches our experimentalists' area of expertise and has ideally been demonstrated in earlier experimental works but without dielectric characterization.
    \item[strong candidate]: promising in some ways, i.e. high calculated band gap and permittivity but perhaps no existing literature reporting successful prior synthesis or compound looks challenging to make (e.g. might require aerobic environment)
    \item[weak candidate]: less promising in terms of simulated properties but potentially easier to make than other materials with superior expected properties
    \item[discarded]: failures of our screening method, usually due to existing literature indicating properties are not as we predict such as when a material was previously synthesized but reported as black, indicating a small band gap.
\end{description}

This interactive selection tool proved invaluable for extracting maximum utility and insight from the data we generated and resulted in identifying two candidates for final selection as suitable candidates for experimental synthesis.

We use GitHub Pages to host a figure-only version of this web interface at \href{https://janosh.github.io/dielectrics}{janosh.github.io/dielectrics}. It is set up with continuous integration to update automatically as new data is generated. It has no write access to the database and hence cannot be used to annotate or categorize candidate materials but serves as a user-friendly public entry point to the most promising results in our database that requires no setup nor technical knowledge to use.

\subsection{Synthesis Details}
\label{sec:synthesis-details}

\CsTaTeO{} was synthesized using standard solid-state synthesis techniques. Stoichiometric amounts of Cs2CO3 (Alfa Aesar, 99.95), \ch{Ta2O5} (Afla Asear, 99.999), and Te(OH)6 (Aldrich, 99.5) were added to an agate pestle and mortar and ground to homogenize the precursors before calcining at \SI{400}{\celsius} for \SI{24}{h} in an \ch{Al2O3} crucible. After calcining, samples were reground and pressed into a \SI{10}{mm} disk and annealed at \SI{750}{\celsius} for \SI{48}{h} in a covered \ch{Al2O3} crucible to form the final product.

\BiZrO{} was synthesized using an ethylenediaminetetraacetic acid (EDTA) and nitrate chelation and combustion, similar to a sol-gel process, a reaction modified from \cite{pandey_metastable_2018}. Equal molar quantities of \ch{Bi2O3} and \ch{ZrO(NO3)2} were added to separate beakers and dissolved in minimal amounts of concentrated nitric acid by stirring with a magnetic stir bar. Once dissolved, these two solutions were mixed with a 4 times molar excess of EDTA to ensure chelation. The solution was then heated at \SI{80}{\celsius} until all liquid evaporated, leaving a brownish-white powder that was amorphous to X-rays. The amorphous powder was then calcined as a loose powder in a furnace inside a \ch{Al2O3} crucible at temperatures from \SIrange{550}{750}{\celsius} in \SI{50}{\celsius} increments for \SI{1}{h}. The samples heated at \SI{650}{\celsius} produced the sharpest XRD peaks, without any trace of impurities. Samples heated higher than \SI{650}{\celsius} or for longer than \SI{1}{h} resulted in the decomposition of the sample into \ch{Bi2O3} and \ch{ZrO2}, indicating metastability.

Both samples were sintered using spark plasma sintering. Pure phase samples were loaded into \SI{10}{mm} graphite dies in a Thermal Technology LLC DCS10 furnace. Samples ($\sim$\SI{0.75}{g}) were loaded into a \SI{10}{mm} diameter graphite die lined with a graphite foil and loaded into a sample chamber which was evacuated and backfilled with \ch{He} three times. The sample was pressed uniaxially at \SI{60}{MPa}, heated to the desired temperature at a rate of \SI{200}{C/min}, held for \SI{1}{min}, and cooled at the same rate. The \CsTaTeO{} sample was heated to a maximum temperature of \SI{750}{\celsius} and the \BiZrO{} sample was heated to \SI{600}{\celsius} resulting in samples with 92\% and 94\% of theoretical densities, respectively.

Diffuse reflectance measurements were taken on powdered using a Cary 5000 UV–Vis–NIR Spectrometer. Dielectric permittivity data was collected on sintered samples that had been thinned to a thickness of \SI{1}{mm} and sputtered with gold electrodes. Data was collected using an Agilent 4980A instrument with a home-built sample holder and a program created in LABVIEW. X-ray diffraction data was collected using a Paralytical X'pert Pro instrument with \ch{Co} K$\alpha$1 ($\lambda = \SI{1.788960}{\A}$) radiation. Rietveld analysis was carried out using Topas Academic on these X-ray data. Initial refinements started with parameters identified using the Pawley method. Final refinements included lattice parameters, atomic positions, atomic displacement parameters, profile parameters and the background.

\section{Discussion}
\label{sec:discussion}

We have demonstrated a high-throughput workflow for dielectric materials discovery that combines data-driven and first-principles methods.
We show in \cref{tab:hit-rate-comparison} that this combination achieves improved enrichment of high $\fom$ materials than ab-initio methods alone.

By deploying this workflow into practice, we identified and synthesized two candidate materials, \CsTaTeO{} and \BiZrO{}. After careful Rietveld analysis to verify we realized the target structures, we measured their band gaps and dielectric properties.
\BiZrO{} shows strong promise for electronic applications given its measured band gap of \SI{2.27}{eV}, dielectric constant of 20.5 and its relatively available constituent elements.
\CsTaTeO{} is a black semiconductor with a low band gap of \SI{1.05}{eV} and dielectric constant of 26, making it unsuitable for electronic applications.
However, we emphasize this structure was generated via element substitution by our workflow with no prior reports in the ICSD or MP.
We thus demonstrated successful de novo synthesis on a challenging metastable phase and established a prior for the dielectric properties of similar materials in this largely unexplored region of chemical space.
This outcome shows that ML-driven thermodynamic stability prediction has matured enough in reliability to be effectively incorporated into a complex multi-step workflow.
This requires sufficient trust in the method to attempt a risky metastable synthesis in an unknown chemical system.

The biggest failure mode in our funnel search was the weakness of our band gap ML model.
It incurred a high false-positive rate, predicting many generated metallic structures as semiconductors or insulators.
Although there is significant room for improvement in ML band gap prediction, it was not the main focus of this work.
We consider accurate band gap models to be an unsolved problem in materials informatics and encourage more efforts be directed at it.
Models that predict a spectrum rather than a single scalar may be an interesting avenue to pursue.
Predicting the electronic density of states (eDOS) like Mat2Spec \cite{kong_density_2021} and inferring the band gap from that also opens the door to more nuanced loss functions and increased regularization during training.
Sufficiently complex models with good inductive bias may learn more subtle trends from this approach.
It should be noted, however, that Mat2Spec refrained from reporting band gaps inferred from their eDOS predictions, potentially indicating more work is required to unlock such benefits.
\citeauthor*{shoghi_molecules_2023} \cite{shoghi_molecules_2023} is a more recent work demonstrating impressive band gap accuracy on the matbench MP $\egap$ task after pre-training on many large but non-cognate materials prediction tasks.
This suggests that perhaps current model architectures and training methods can be sufficient.
Achieving reliable ML band gap prediction could be a matter of careful data curation and model pre-training.

However, the challenge of predicting band gaps in our workflow is not restricted to ML but carries through to DFT.
PBE exhibits an unusual severe overestimation ($\egap^\text{PBE} > \SI{2.09}{eV}$) of the experimental band gap of ($\egap^\text{exp} = \SI{1.05}{eV}$) of \CsTaTeO{}.
Although defect chemistry may play a role in this effect, there are obvious computing limitations in a high-throughput workflow, making the simulation of defect effects cost-prohibitive.
One obvious improvement to narrow the gap between simulation and reality is to employ higher levels of theory such as r2SCAN or even to incorporate a third computational filter to the funnel in the form of hybrid functionals such as HSE, applied sparingly to compounds that have passed ML and PBE filters but before attempting experimental synthesis.

\section*{Code and Data Availability}
\label{sec:code-availability}

The MIT-licensed code for this work can be found at \url{https://github.com/janosh/dielectrics} and as a Zenodo archive at \href{https://doi.org/10.5281/zenodo.10456384}{https://doi.org/10.5281/zenodo.10456384}.
Zenodo includes a complete dump of our DFPT dataset.
Our live data is also publicly accessible through a MongoDB M2 Atlas instance with the schema of an \texttt{atomate} \texttt{tasks} collection.
It can be queried free of charge and without registration using the read-only database credentials and example code snippet provided in the GitHub readme.
This requires \href{https://pymongo.readthedocs.io}{\texttt{pymongo}} or any other MongoDB language driver.
The query syntax will be familiar to users of the (legacy) Materials Project \texttt{MPRester} API.
We used Materials Project data from the (\href{https://docs.materialsproject.org/changes/database-versions}{v2020.09.08}) database release and a cleaned version of the WBM dataset \cite{wang_predicting_2021} available at \url{https://figshare.com/articles/dataset/22715158}.

\printbibliography

@article{bedard_reconfigurable_2018,
  title = {Reconfigurable System for Automated Optimization of Diverse Chemical Reactions},
  author = {Bédard, Anne-Catherine and Adamo, Andrea and Aroh, Kosi C. and Russell, M. Grace and Bedermann, Aaron A. and Torosian, Jeremy and Yue, Brian and Jensen, Klavs F. and Jamison, Timothy F.},
  date = {2018-09-21},
  journaltitle = {Science},
  volume = {361},
  number = {6408},
  pages = {1220--1225},
  publisher = {{American Association for the Advancement of Science}},
  doi = {10.1126/science.aat0650},
  url = {https://www.science.org/doi/full/10.1126/science.aat0650},
  urldate = {2024-01-03}
}

@article{bergerhoff_inorganic_1983,
  title = {The Inorganic Crystal Structure Data Base},
  author = {Bergerhoff, G. and Hundt, R. and Sievers, R. and Brown, I. D.},
  date = {1983-05-01},
  journaltitle = {Journal of Chemical Information and Computer Sciences},
  shortjournal = {J. Chem. Inf. Comput. Sci.},
  volume = {23},
  number = {2},
  pages = {66--69},
  publisher = {{American Chemical Society}},
  issn = {0095-2338},
  doi = {10.1021/ci00038a003},
  url = {https://pubs.acs.org/doi/abs/10.1021/ci00038a003},
  urldate = {2020-08-31}
}

@article{blochl_projector_1994,
  title = {Projector Augmented-Wave Method},
  author = {Blöchl, P. E.},
  date = {1994-12-15},
  journaltitle = {Physical Review B},
  shortjournal = {Phys. Rev. B},
  volume = {50},
  number = {24},
  pages = {17953--17979},
  publisher = {{American Physical Society}},
  doi = {10.1103/PhysRevB.50.17953},
  url = {https://link.aps.org/doi/10.1103/PhysRevB.50.17953},
  urldate = {2022-02-11}
}

@article{burger_mobile_2020,
  title = {A Mobile Robotic Chemist},
  author = {Burger, Benjamin and Maffettone, Phillip M. and Gusev, Vladimir V. and Aitchison, Catherine M. and Bai, Yang and Wang, Xiaoyan and Li, Xiaobo and Alston, Ben M. and Li, Buyi and Clowes, Rob and Rankin, Nicola and Harris, Brandon and Sprick, Reiner Sebastian and Cooper, Andrew I.},
  date = {2020-07},
  journaltitle = {Nature},
  volume = {583},
  number = {7815},
  pages = {237--241},
  publisher = {{Nature Publishing Group}},
  issn = {1476-4687},
  doi = {10.1038/s41586-020-2442-2},
  url = {https://www.nature.com/articles/s41586-020-2442-2},
  urldate = {2024-01-03},
  issue = {7815}
}

@article{chiu_investigations_2022,
  title = {Investigations of Mechanical Properties and Deformation Behaviors of the {{Cr}} Modified {{Ti}}–{{Au}} Shape Memory Alloys},
  author = {Chiu, Wan–Ting and Wakabayashi, Kaoru and Umise, Akira and Tahara, Masaki and Inamura, Tomonari and Hosoda, Hideki},
  date = {2022-03-15},
  journaltitle = {Journal of Alloys and Compounds},
  shortjournal = {Journal of Alloys and Compounds},
  volume = {897},
  pages = {163134},
  issn = {0925-8388},
  doi = {10.1016/j.jallcom.2021.163134},
  url = {https://www.sciencedirect.com/science/article/pii/S0925838821045448},
  urldate = {2023-12-30}
}

@article{choi_development_2011,
  title = {Development of Hafnium Based High-k Materials—{{A}} Review},
  author = {Choi, J.H. and Mao, Y. and Chang, J.P.},
  date = {2011-07},
  journaltitle = {Materials Science and Engineering: R: Reports},
  shortjournal = {Materials Science and Engineering: R: Reports},
  volume = {72},
  number = {6},
  pages = {97--136},
  issn = {0927796X},
  doi = {10.1016/j.mser.2010.12.001},
  url = {https://linkinghub.elsevier.com/retrieve/pii/S0927796X10001683},
  urldate = {2023-11-30}
}

@article{choudhary_highthroughput_2020,
  title = {High-Throughput Density Functional Perturbation Theory and Machine Learning Predictions of Infrared, Piezoelectric, and Dielectric Responses},
  author = {Choudhary, Kamal and Garrity, Kevin F. and Sharma, Vinit and Biacchi, Adam J. and Hight Walker, Angela R. and Tavazza, Francesca},
  date = {2020-05-27},
  journaltitle = {npj Computational Materials},
  shortjournal = {npj Comput Mater},
  volume = {6},
  number = {1},
  pages = {1--13},
  publisher = {{Nature Publishing Group}},
  issn = {2057-3960},
  doi = {10.1038/s41524-020-0337-2},
  url = {https://www.nature.com/articles/s41524-020-0337-2},
  urldate = {2022-06-01},
  issue = {1}
}

@article{davies_computational_2016,
  title = {Computational {{Screening}} of {{All Stoichiometric Inorganic Materials}}},
  author = {Davies, Daniel W. and Butler, Keith T. and Jackson, Adam J. and Morris, Andrew and Frost, Jarvist M. and Skelton, Jonathan M. and Walsh, Aron},
  date = {2016-10-13},
  journaltitle = {Chem},
  shortjournal = {Chem},
  volume = {1},
  number = {4},
  eprint = {27790643},
  eprinttype = {pmid},
  pages = {617--627},
  issn = {2451-9294},
  doi = {10.1016/j.chempr.2016.09.010},
  pmcid = {PMC5074417}
}

@online{fedorova_writes_2022,
  title = {Writes {{Hurt}}: {{Lessons}} in {{Cache Design}} for {{Optane NVRAM}}},
  shorttitle = {Writes {{Hurt}}},
  author = {Fedorova, Alexandra and Smith, Keith and Bostic, Keith and Gorrod, Alexander and LoVerso, Sue and Cahill, Michael},
  date = {2022-05-24},
  eprint = {2205.14122},
  eprinttype = {arxiv},
  eprintclass = {cs},
  doi = {10.48550/arXiv.2205.14122},
  url = {http://arxiv.org/abs/2205.14122},
  urldate = {2023-10-16},
  pubstate = {preprint}
}

@article{feng_unraveling_2021,
  title = {Unraveling the {{Principles}} of {{Lattice Disorder Degree}} of {{Bi2 B2 O7}} ({{B}} = {{Sn}}, {{Ti}}, {{Zr}}) {{Compounds}} on {{Activating Gas Phase O2}} for {{Soot Combustion}}},
  author = {Feng, Xiaohui and Xu, Junwei and Xu, Xianglan and Zhang, Shijing and Ma, Jun and Fang, Xiuzhong and Wang, Xiang},
  date = {2021-10-01},
  journaltitle = {ACS Catalysis},
  shortjournal = {ACS Catal.},
  volume = {11},
  number = {19},
  pages = {12112--12122},
  issn = {2155-5435, 2155-5435},
  doi = {10.1021/acscatal.1c03075},
  url = {https://pubs.acs.org/doi/10.1021/acscatal.1c03075},
  urldate = {2023-11-30}
}

@article{fukina_crystal_2019,
  title = {Crystal Structure and Thermal Behavior of Pyrochlores {{CsTeMoO6}} and {{RbTe1}}.{{25Mo0}}.{{75O6}}},
  author = {Fukina, D.G. and Suleimanov, E.V. and Fukin, G.K. and Boryakov, A.V. and Protasova, S.G. and Ionov, A.M. and Guseinov, D.V. and Istomin, L.A.},
  date = {2019-04},
  journaltitle = {Journal of Solid State Chemistry},
  shortjournal = {Journal of Solid State Chemistry},
  volume = {272},
  pages = {47--54},
  issn = {00224596},
  doi = {10.1016/j.jssc.2019.01.026},
  url = {https://linkinghub.elsevier.com/retrieve/pii/S0022459619300398},
  urldate = {2023-11-30}
}

@article{fukina_structure_2021,
  title = {Structure Analysis and Electronic Properties of {{ATe4}}+0.{{5Te6}}+1.5-{{xM6}}+{{xO6}} ({{A}}={{Rb}}, {{Cs}}, {{M6}}+={{Mo}}, {{W}}) Solid Solutions with Beta-Pyrochlore Structure},
  author = {Fukina, Diana G. and Suleimanov, Eugeny V. and Boryakov, Aleksey V. and Zubkov, Sergey Yu and Koryagin, Andrey V. and Volkova, Natalia S. and Gorshkov, Alexey P.},
  date = {2021-01},
  journaltitle = {Journal of Solid State Chemistry},
  shortjournal = {Journal of Solid State Chemistry},
  volume = {293},
  pages = {121787},
  issn = {00224596},
  doi = {10.1016/j.jssc.2020.121787},
  url = {https://linkinghub.elsevier.com/retrieve/pii/S0022459620306186},
  urldate = {2023-11-30}
}

@article{galati_cation_2008,
  title = {Cation Displacements and the Structures of the Superconducting Pyrochlore Osmates {{A Os}} 2 {{O}} 6 ( {{A}} = {{K}} , {{Rb}}, and {{Cs}})},
  author = {Galati, Rosa and Simon, Charles and Henry, Paul F. and Weller, Mark T.},
  date = {2008-03-21},
  journaltitle = {Physical Review B},
  shortjournal = {Phys. Rev. B},
  volume = {77},
  number = {10},
  pages = {104523},
  issn = {1098-0121, 1550-235X},
  doi = {10.1103/PhysRevB.77.104523},
  url = {https://link.aps.org/doi/10.1103/PhysRevB.77.104523},
  urldate = {2023-11-30}
}

@article{gaultois_perspective_2016,
  title = {Perspective: {{Web-based}} Machine Learning Models for Real-Time Screening of Thermoelectric Materials Properties},
  shorttitle = {Perspective},
  author = {Gaultois, Michael W. and Oliynyk, Anton O. and Mar, Arthur and Sparks, Taylor D. and Mulholland, Gregory J. and Meredig, Bryce},
  date = {2016-05-01},
  journaltitle = {APL Materials},
  shortjournal = {APL Materials},
  volume = {4},
  number = {5},
  pages = {053213},
  publisher = {{American Institute of Physics}},
  doi = {10.1063/1.4952607},
  url = {https://aip.scitation.org/doi/10.1063/1.4952607},
  urldate = {2020-08-31}
}

@article{glawe_optimal_2016,
  title = {The Optimal One Dimensional Periodic Table: A Modified {{Pettifor}} Chemical Scale from Data Mining},
  shorttitle = {The Optimal One Dimensional Periodic Table},
  author = {Glawe, Henning and Sanna, Antonio and Gross, E. K. U. and Marques, Miguel A. L.},
  date = {2016-09},
  journaltitle = {New Journal of Physics},
  shortjournal = {New J. Phys.},
  volume = {18},
  number = {9},
  pages = {093011},
  publisher = {{IOP Publishing}},
  issn = {1367-2630},
  doi = {10.1088/1367-2630/18/9/093011},
  url = {https://doi.org/10.1088/1367-2630/18/9/093011},
  urldate = {2022-02-09}
}

@article{gonze_dynamical_1997,
  title = {Dynamical Matrices, {{Born}} Effective Charges, Dielectric Permittivity Tensors, and Interatomic Force Constants from Density-Functional Perturbation Theory},
  author = {Gonze, Xavier and Lee, Changyol},
  date = {1997-04-15},
  journaltitle = {Physical Review B},
  shortjournal = {Phys. Rev. B},
  volume = {55},
  number = {16},
  pages = {10355--10368},
  publisher = {{American Physical Society}},
  doi = {10.1103/PhysRevB.55.10355},
  url = {https://link.aps.org/doi/10.1103/PhysRevB.55.10355},
  urldate = {2023-10-01}
}

@unpublished{goodall_rapid_2021,
  title = {Rapid {{Discovery}} of {{Novel Materials}} by {{Coordinate-free Coarse Graining}}},
  author = {Goodall, Rhys E. A. and Parackal, Abhijith S. and Faber, Felix A. and Armiento, Rickard and Lee, Alpha A.},
  date = {2021-10-16},
  eprint = {2106.11132},
  eprinttype = {arxiv},
  eprintclass = {cond-mat, physics:physics},
  url = {http://arxiv.org/abs/2106.11132},
  urldate = {2022-02-09}
}

@article{guan_bimetallic_2021,
  title = {Bimetallic Monolayer Catalyst Breaks the Activity–Selectivity Trade-off on Metal Particle Size for Efficient Chemoselective Hydrogenations},
  author = {Guan, Qiaoqiao and Zhu, Chuwei and Lin, Yue and Vovk, Evgeny I. and Zhou, Xiaohong and Yang, Yong and Yu, Hancheng and Cao, Lina and Wang, Hengwei and Zhang, Xiaohui and Liu, Xinyu and Zhang, Mengkai and Wei, Shiqiang and Li, Wei-Xue and Lu, Junling},
  date = {2021-10},
  journaltitle = {Nature Catalysis},
  shortjournal = {Nat Catal},
  volume = {4},
  number = {10},
  pages = {840--849},
  publisher = {{Nature Publishing Group}},
  issn = {2520-1158},
  doi = {10.1038/s41929-021-00679-x},
  url = {https://www.nature.com/articles/s41929-021-00679-x},
  urldate = {2023-12-30},
  issue = {10}
}

@unpublished{he_deep_2015,
  title = {Deep {{Residual Learning}} for {{Image Recognition}}},
  author = {He, Kaiming and Zhang, Xiangyu and Ren, Shaoqing and Sun, Jian},
  date = {2015-12-10},
  eprint = {1512.03385},
  eprinttype = {arxiv},
  eprintclass = {cs},
  url = {http://arxiv.org/abs/1512.03385},
  urldate = {2020-08-05}
}

@article{hinkle_novel_2004,
  title = {A Novel Approach for Determining the Effective Tunneling Mass of Electrons in {{HfO2}} and Other High-{{K}} Alternative Gate Dielectrics for Advanced {{CMOS}} Devices},
  author = {Hinkle, C. L and Fulton, C and Nemanich, R. J and Lucovsky, G},
  date = {2004-04-01},
  journaltitle = {Microelectronic Engineering},
  shortjournal = {Microelectronic Engineering},
  series = {Proceedings of the 13th {{Biennial Conference}} on {{Insulating Films}} on {{Semiconductors}}},
  volume = {72},
  number = {1},
  pages = {257--262},
  issn = {0167-9317},
  doi = {10.1016/j.mee.2003.12.047},
  url = {https://www.sciencedirect.com/science/article/pii/S0167931703006191},
  urldate = {2022-06-01}
}

@online{horton_crystal_2023,
  title = {Crystal {{Toolkit}}: {{A Web App Framework}} to {{Improve Usability}} and {{Accessibility}} of {{Materials Science Research Algorithms}}},
  shorttitle = {Crystal {{Toolkit}}},
  author = {Horton, Matthew and Shen, Jimmy-Xuan and Burns, Jordan and Cohen, Orion and Chabbey, François and Ganose, Alex M. and Guha, Rishabh and Huck, Patrick and Li, Hamming Howard and McDermott, Matthew and Montoya, Joseph and Moore, Guy and Munro, Jason and O'Donnell, Cody and Ophus, Colin and Petretto, Guido and Riebesell, Janosh and Wetizner, Steven and Wander, Brook and Winston, Donald and Yang, Ruoxi and Zeltmann, Steven and Jain, Anubhav and Persson, Kristin A.},
  date = {2023-02-27},
  eprint = {2302.06147},
  eprinttype = {arxiv},
  eprintclass = {cond-mat},
  doi = {10.48550/arXiv.2302.06147},
  url = {http://arxiv.org/abs/2302.06147},
  urldate = {2023-02-16},
  pubstate = {preprint}
}

@article{hossain_visualization_2019,
  title = {Visualization of {{Bioinformatics Data}} with {{Dash Bio}}},
  author = {Hossain, Shammamah},
  date = {2019},
  journaltitle = {Proceedings of the 18th Python in Science Conference},
  pages = {126--133},
  doi = {10.25080/Majora-7ddc1dd1-012},
  url = {https://conference.scipy.org/proceedings/scipy2019/shammamah_hossain.html},
  urldate = {2023-10-16},
  eventtitle = {Proceedings of the 18th {{Python}} in {{Science Conference}}}
}

@article{jain_commentary_2013,
  title = {Commentary: {{The Materials Project}}: {{A}} Materials Genome Approach to Accelerating Materials Innovation},
  shorttitle = {Commentary},
  author = {Jain, Anubhav and Ong, Shyue Ping and Hautier, Geoffroy and Chen, Wei and Richards, William Davidson and Dacek, Stephen and Cholia, Shreyas and Gunter, Dan and Skinner, David and Ceder, Gerbrand and Persson, Kristin A.},
  date = {2013-07},
  journaltitle = {APL Materials},
  volume = {1},
  number = {1},
  pages = {011002},
  publisher = {{American Institute of Physics}},
  doi = {10.1063/1.4812323},
  url = {https://aip.scitation.org/doi/10.1063%2F1.4812323},
  urldate = {2022-02-21}
}

@article{jain_fireworks_2015,
  title = {{{FireWorks}}: A Dynamic Workflow System Designed for High-Throughput Applications},
  shorttitle = {{{FireWorks}}},
  author = {Jain, Anubhav and Ong, Shyue Ping and Chen, Wei and Medasani, Bharat and Qu, Xiaohui and Kocher, Michael and Brafman, Miriam and Petretto, Guido and Rignanese, Gian-Marco and Hautier, Geoffroy and Gunter, Daniel and Persson, Kristin A.},
  date = {2015},
  journaltitle = {Concurrency and Computation: Practice and Experience},
  volume = {27},
  number = {17},
  pages = {5037--5059},
  issn = {1532-0634},
  doi = {10.1002/cpe.3505},
  url = {https://onlinelibrary.wiley.com/doi/abs/10.1002/cpe.3505},
  urldate = {2022-02-21}
}

@article{jayaraman_bridging_2020,
  title = {Bridging and Synergistic Effect of the Pyrochlore like {{Bi2 Zr2 O7}} Structure with Robust {{CdCuS}} Solid Solution for Durable Photocatalytic Removal of the Organic Pollutants},
  author = {Jayaraman, Venkatesan and Ayappan, Chinnadurai and Palanivel, Baskaran and Mani, Alagiri},
  date = {2020},
  journaltitle = {RSC Advances},
  shortjournal = {RSC Adv.},
  volume = {10},
  number = {15},
  pages = {8880--8894},
  issn = {2046-2069},
  doi = {10.1039/D0RA00644K},
  url = {http://xlink.rsc.org/?DOI=D0RA00644K},
  urldate = {2023-11-30}
}

@article{king_rise_2011,
  title = {Rise of the {{Robo Scientists}}},
  author = {King, Ross D.},
  date = {2011},
  journaltitle = {Scientific American},
  volume = {304},
  number = {1},
  eprint = {26002355},
  eprinttype = {jstor},
  pages = {72--77},
  publisher = {{Scientific American, a division of Nature America, Inc.}},
  issn = {0036-8733},
  url = {https://www.jstor.org/stable/26002355},
  urldate = {2024-01-03}
}

@unpublished{kong_density_2021,
  title = {Density of {{States Prediction}} for {{Materials Discovery}} via {{Contrastive Learning}} from {{Probabilistic Embeddings}}},
  author = {Kong, Shufeng and Ricci, Francesco and Guevarra, Dan and Neaton, Jeffrey B. and Gomes, Carla P. and Gregoire, John M.},
  date = {2021-10-21},
  eprint = {2110.11444},
  eprinttype = {arxiv},
  eprintclass = {cond-mat, physics:physics},
  url = {http://arxiv.org/abs/2110.11444},
  urldate = {2021-10-28}
}

@article{kresse_efficiency_1996,
  title = {Efficiency of Ab-Initio Total Energy Calculations for Metals and Semiconductors Using a Plane-Wave Basis Set},
  author = {Kresse, Georg and Furthmüller, Jürgen},
  date = {1996},
  journaltitle = {Computational materials science},
  volume = {6},
  number = {1},
  pages = {15--50}
}

@article{kresse_efficient_1996,
  title = {Efficient Iterative Schemes for Ab Initio Total-Energy Calculations Using a Plane-Wave Basis Set},
  author = {Kresse, G. and Furthmüller, J.},
  date = {1996-10-15},
  journaltitle = {Physical Review B},
  shortjournal = {Phys. Rev. B},
  volume = {54},
  number = {16},
  pages = {11169--11186},
  doi = {10.1103/PhysRevB.54.11169},
  url = {https://link.aps.org/doi/10.1103/PhysRevB.54.11169},
  urldate = {2019-08-22}
}

@article{kubelka_article_1931,
  title = {An {{Article}} on {{Optics}} of {{Paint Layers}}},
  author = {Kubelka, Paul and Munk, Franz},
  date = {1931-09},
  journaltitle = {Z. Tech. Phys},
  number = {12},
  pages = {593--601}
}

@article{kurlla_greenengineered_2023,
  title = {Green-Engineered Synthesis of {{Bi2Zr2O7 NPs}}: Excellent Performance on Electrochemical Sensor and Sunlight-Driven Photocatalytic Studies},
  shorttitle = {Green-Engineered Synthesis of {{Bi2Zr2O7 NPs}}},
  author = {Kurlla, Pompapathi and Shivram, Anantharaju Kurupalya and Kottam, Nagaraju and Siddegowda, Surendra Boppanahalli and Subramaniam, Meena and Bogegowda, Uma and Subramanya, Malini and Chowdhury, Arpita Paul and Narasimhan, Renuka Lakshmi},
  date = {2023-01-05},
  journaltitle = {Environmental Science and Pollution Research},
  shortjournal = {Environ Sci Pollut Res},
  issn = {1614-7499},
  doi = {10.1007/s11356-022-24760-5},
  url = {https://link.springer.com/10.1007/s11356-022-24760-5},
  urldate = {2023-11-30}
}

@unpublished{lakshminarayanan_simple_2016,
  title = {Simple and {{Scalable Predictive Uncertainty Estimation}} Using {{Deep Ensembles}}},
  author = {Lakshminarayanan, Balaji and Pritzel, Alexander and Blundell, Charles},
  date = {2016-12-05},
  eprint = {1612.01474},
  eprinttype = {arxiv},
  eprintclass = {cs, stat},
  url = {http://arxiv.org/abs/1612.01474},
  urldate = {2019-03-03}
}

@article{langreth_beyond_1983,
  title = {Beyond the Local-Density Approximation in Calculations of Ground-State Electronic Properties},
  author = {Langreth, David C. and Mehl, M. J.},
  date = {1983-08-15},
  journaltitle = {Physical Review B},
  shortjournal = {Phys. Rev. B},
  volume = {28},
  number = {4},
  pages = {1809--1834},
  publisher = {{American Physical Society}},
  doi = {10.1103/PhysRevB.28.1809},
  url = {https://link.aps.org/doi/10.1103/PhysRevB.28.1809},
  urldate = {2022-02-11}
}

@article{lee_comparative_2007,
  title = {Comparative Study of Electronic Structures and Dielectric Properties of Alumina Polymorphs by First-Principles Methods},
  author = {Lee, Choong-Ki and Cho, Eunae and Lee, Hyo-Sug and Seol, Kwang Soo and Han, Seungwu},
  date = {2007-12-10},
  journaltitle = {Physical Review B},
  shortjournal = {Phys. Rev. B},
  volume = {76},
  number = {24},
  pages = {245110},
  publisher = {{American Physical Society}},
  doi = {10.1103/PhysRevB.76.245110},
  url = {https://link.aps.org/doi/10.1103/PhysRevB.76.245110},
  urldate = {2022-03-26}
}

@article{li_effect_2021,
  title = {Effect of Shape Memory Alloys on the Mechanical Properties of Metallic Glasses: {{A}} Molecular Dynamics Study},
  shorttitle = {Effect of Shape Memory Alloys on the Mechanical Properties of Metallic Glasses},
  author = {Li, W. W. and Song, H. Y. and Dai, J. L. and Wang, J. Y. and An, M. R. and Li, Y. L.},
  date = {2021-02-01},
  journaltitle = {Computational Materials Science},
  shortjournal = {Computational Materials Science},
  volume = {187},
  pages = {110088},
  issn = {0927-0256},
  doi = {10.1016/j.commatsci.2020.110088},
  url = {https://www.sciencedirect.com/science/article/pii/S0927025620305796},
  urldate = {2023-12-30}
}

@article{liu_bi2zr2o7_2018,
  title = {{{Bi2Zr2O7}} Nanoparticles Synthesized by Soft-Templated Sol-Gel Methods for Visible-Light-Driven Catalytic Degradation of Tetracycline},
  author = {Liu, Xiaowei and Huang, Lihui and Wu, Xueyuan and Wang, Zexiang and Dong, Guihua and Wang, Chuang and Liu, Yangyang and Wang, Lisha},
  date = {2018-11},
  journaltitle = {Chemosphere},
  shortjournal = {Chemosphere},
  volume = {210},
  pages = {424--432},
  issn = {00456535},
  doi = {10.1016/j.chemosphere.2018.07.040},
  url = {https://linkinghub.elsevier.com/retrieve/pii/S0045653518312980},
  urldate = {2023-11-30}
}

@article{liutkova_ca_2023,
  title = {Ca/{{ZSM-5}} Catalysts for the Methanol-to-Hydrocarbons Reaction: {{Activity}} – {{Selectivity}} Trade-Off?},
  shorttitle = {Ca/{{ZSM-5}} Catalysts for the Methanol-to-Hydrocarbons Reaction},
  author = {Liutkova, Anna and Kosinov, Nikolay and Hensen, Emiel J. M.},
  date = {2023-12-01},
  journaltitle = {Journal of Catalysis},
  shortjournal = {Journal of Catalysis},
  volume = {428},
  pages = {115169},
  issn = {0021-9517},
  doi = {10.1016/j.jcat.2023.115169},
  url = {https://www.sciencedirect.com/science/article/pii/S0021951723004141},
  urldate = {2023-12-30}
}

@article{lunt_modular_2024,
  title = {Modular, {{Multi-Robot Integration}} of {{Laboratories}}: {{An Autonomous Workflow}} for {{Solid-State Chemistry}}},
  shorttitle = {Modular, {{Multi-Robot Integration}} of {{Laboratories}}},
  author = {Lunt, Amy and Fakhruldeen, Hatem and Pizzuto, Gabriella and Longley, Louis and White, Alex and Rankin, Nicola and Clowes, Rob and Alston, Ben and Gigli, Lucia and Day, Graeme Matthew},
  date = {2024},
  journaltitle = {Chemical Science},
  publisher = {{Royal Society of Chemistry}},
  url = {https://pubs.rsc.org/en/content/articlehtml/2024/sc/d3sc06206f},
  urldate = {2024-01-04}
}

@article{luo_new_2018,
  title = {A New Approach to Preparing {{Bi}} {\textsubscript{2}} {{Zr}} {\textsubscript{2}} {{O}} {\textsubscript{7}} Photocatalysts for Dye Degradation},
  author = {Luo, Yijia and Cao, Liyun and Huang, Jianfeng and Feng, Liangliang and Yao, Chunyan},
  date = {2018-01-18},
  journaltitle = {Materials Research Express},
  shortjournal = {Mater. Res. Express},
  volume = {5},
  number = {1},
  pages = {015039},
  issn = {2053-1591},
  doi = {10.1088/2053-1591/aaa584},
  url = {https://iopscience.iop.org/article/10.1088/2053-1591/aaa584},
  urldate = {2023-11-30}
}

@article{luo_synthesis_2019,
  title = {Synthesis, Characterization and Photocatalytic Properties of Nanoscale Pyrochlore Type {{Bi2Zr2O7}}},
  author = {Luo, Yijia and Cao, Liyun and Feng, Liangliang and Huang, Jianfeng and Yang, Liuqing and Yao, Chunyan and Cheng, Yayi},
  date = {2019-01},
  journaltitle = {Materials Science and Engineering: B},
  shortjournal = {Materials Science and Engineering: B},
  volume = {240},
  pages = {133--139},
  issn = {09215107},
  doi = {10.1016/j.mseb.2019.01.017},
  url = {https://linkinghub.elsevier.com/retrieve/pii/S0921510719300236},
  urldate = {2023-11-30}
}

@article{mathew_atomate_2017,
  title = {Atomate: {{A}} High-Level Interface to Generate, Execute, and Analyze Computational Materials Science Workflows},
  shorttitle = {Atomate},
  author = {Mathew, Kiran and Montoya, Joseph H. and Faghaninia, Alireza and Dwarakanath, Shyam and Aykol, Muratahan and Tang, Hanmei and Chu, Iek-heng and Smidt, Tess and Bocklund, Brandon and Horton, Matthew and Dagdelen, John and Wood, Brandon and Liu, Zi-Kui and Neaton, Jeffrey and Ong, Shyue Ping and Persson, Kristin and Jain, Anubhav},
  date = {2017-11-01},
  journaltitle = {Computational Materials Science},
  shortjournal = {Computational Materials Science},
  volume = {139},
  pages = {140--152},
  issn = {0927-0256},
  doi = {10.1016/j.commatsci.2017.07.030},
  url = {https://www.sciencedirect.com/science/article/pii/S0927025617303919},
  urldate = {2022-02-21}
}

@inproceedings{nair_rectified_2010,
  title = {Rectified {{Linear Units Improve Restricted Boltzmann Machines}}},
  author = {Nair, Vinod and Hinton, Geoffrey E.},
  date = {2010-01-01},
  url = {https://openreview.net/forum?id=rkb15iZdZB},
  urldate = {2022-02-15},
  eventtitle = {{{ICML}}}
}

@article{ong_python_2013,
  title = {Python {{Materials Genomics}} (Pymatgen): {{A}} Robust, Open-Source Python Library for Materials Analysis},
  shorttitle = {Python {{Materials Genomics}} (Pymatgen)},
  author = {Ong, Shyue Ping and Richards, William Davidson and Jain, Anubhav and Hautier, Geoffroy and Kocher, Michael and Cholia, Shreyas and Gunter, Dan and Chevrier, Vincent L. and Persson, Kristin A. and Ceder, Gerbrand},
  date = {2013-02-01},
  journaltitle = {Computational Materials Science},
  shortjournal = {Computational Materials Science},
  volume = {68},
  pages = {314--319},
  issn = {0927-0256},
  doi = {10.1016/j.commatsci.2012.10.028},
  url = {https://www.sciencedirect.com/science/article/pii/S0927025612006295},
  urldate = {2022-02-17}
}

@article{pandey_metastable_2018,
  title = {Metastable {{Bi2Zr2O7}} with {{Pyrochlore-like Structure}}: {{Stabilization}}, {{Oxygen Ion Conductivity}}, and {{Catalytic Properties}}},
  shorttitle = {Metastable {{Bi2Zr2O7}} with {{Pyrochlore-like Structure}}},
  author = {Pandey, Jyoti and Shrivastava, Vipul and Nagarajan, Rajamani},
  date = {2018-11-05},
  journaltitle = {Inorganic Chemistry},
  shortjournal = {Inorg. Chem.},
  volume = {57},
  number = {21},
  pages = {13667--13678},
  publisher = {{American Chemical Society}},
  issn = {0020-1669},
  doi = {10.1021/acs.inorgchem.8b02258},
  url = {https://doi.org/10.1021/acs.inorgchem.8b02258},
  urldate = {2023-10-21}
}

@article{perdew_rationale_1996,
  title = {Rationale for Mixing Exact Exchange with Density Functional Approximations},
  author = {Perdew, John P. and Ernzerhof, Matthias and Burke, Kieron},
  date = {1996-12-08},
  journaltitle = {The Journal of Chemical Physics},
  shortjournal = {J. Chem. Phys.},
  volume = {105},
  number = {22},
  pages = {9982--9985},
  publisher = {{American Institute of Physics}},
  issn = {0021-9606},
  doi = {10.1063/1.472933},
  url = {https://aip.scitation.org/doi/10.1063/1.472933},
  urldate = {2022-02-12}
}

@online{persson_materials_2022,
  title = {Materials {{Project}} :: {{About}}},
  author = {Persson, Kristin},
  date = {2022-02-21},
  url = {https://materialsproject.org/about#db-stats},
  urldate = {2022-02-21},
  organization = {{About the Materials Project}}
}

@article{petousis_benchmarking_2016,
  title = {Benchmarking Density Functional Perturbation Theory to Enable High-Throughput Screening of Materials for Dielectric Constant and Refractive Index},
  author = {Petousis, Ioannis and Chen, Wei and Hautier, Geoffroy and Graf, Tanja and Schladt, Thomas D. and Persson, Kristin A. and Prinz, Fritz B.},
  date = {2016-03-31},
  journaltitle = {Physical Review B},
  shortjournal = {Phys. Rev. B},
  volume = {93},
  number = {11},
  pages = {115151},
  publisher = {{American Physical Society}},
  doi = {10.1103/PhysRevB.93.115151},
  url = {https://link.aps.org/doi/10.1103/PhysRevB.93.115151},
  urldate = {2021-10-08}
}

@article{petousis_high-throughput_2017,
  title = {High-Throughput Screening of Inorganic Compounds for the Discovery of Novel Dielectric and Optical Materials},
  author = {Petousis, Ioannis and Mrdjenovich, David and Ballouz, Eric and Liu, Miao and Winston, Donald and Chen, Wei and Graf, Tanja and Schladt, Thomas D. and Persson, Kristin A. and Prinz, Fritz B.},
  date = {2017-01-31},
  journaltitle = {Scientific Data},
  shortjournal = {Sci Data},
  volume = {4},
  number = {1},
  pages = {160134},
  publisher = {{Nature Publishing Group}},
  issn = {2052-4463},
  doi = {10.1038/sdata.2016.134},
  url = {https://www.nature.com/articles/sdata2016134},
  urldate = {2022-02-06},
  issue = {1}
}

@article{petretto_highthroughput_2018,
  title = {High-Throughput Density-Functional Perturbation Theory Phonons for Inorganic Materials},
  author = {Petretto, Guido and Dwaraknath, Shyam and P. C. Miranda, Henrique and Winston, Donald and Giantomassi, Matteo and family=Setten, given=Michiel J., prefix=van, useprefix=true and Gonze, Xavier and Persson, Kristin A. and Hautier, Geoffroy and Rignanese, Gian-Marco},
  date = {2018-05-01},
  journaltitle = {Scientific Data},
  shortjournal = {Sci Data},
  volume = {5},
  number = {1},
  pages = {180065},
  publisher = {{Nature Publishing Group}},
  issn = {2052-4463},
  doi = {10.1038/sdata.2018.65},
  url = {https://www.nature.com/articles/sdata201865},
  urldate = {2023-10-21},
  issue = {1}
}

@article{ponceortiz_highk_2010,
  title = {High-k Organic, Inorganic, and Hybrid Dielectrics for Low-Voltage Organic Field-Effect Transistors},
  author = {Ponce Ortiz, Rocío and Facchetti, Antonio and Marks, Tobin J.},
  date = {2010-01},
  journaltitle = {Chemical Reviews},
  shortjournal = {Chem Rev},
  volume = {110},
  number = {1},
  eprint = {19852443},
  eprinttype = {pmid},
  pages = {205--239},
  issn = {1520-6890},
  doi = {10.1021/cr9001275}
}

@article{qu_high_2020,
  title = {High Dielectric Ternary Oxides from Crystal Structure Prediction and High-Throughput Screening},
  author = {Qu, Jingyu and Zagaceta, David and Zhang, Weiwei and Zhu, Qiang},
  date = {2020-03-06},
  journaltitle = {Scientific Data},
  shortjournal = {Sci Data},
  volume = {7},
  number = {1},
  pages = {81},
  publisher = {{Nature Publishing Group}},
  issn = {2052-4463},
  doi = {10.1038/s41597-020-0418-6},
  url = {https://www.nature.com/articles/s41597-020-0418-6},
  urldate = {2021-11-30},
  issue = {1}
}

@article{rajashekharaiah_nuv_2019,
  title = {{{NUV}} Light-Induced Visible Green Emissions of {{Erbium-doped}} Hierarchical {{Bi2Zr2O7}} Structures},
  author = {Rajashekharaiah, A.S. and Darshan, G.P. and Basavaraj, R.B. and Naik, Yashwanth V. and Kavyashree, D. and Sharma, S.C. and Nagabhushana, H.},
  date = {2019-09},
  journaltitle = {Optical Materials},
  shortjournal = {Optical Materials},
  volume = {95},
  pages = {109237},
  issn = {09253467},
  doi = {10.1016/j.optmat.2019.109237},
  url = {https://linkinghub.elsevier.com/retrieve/pii/S0925346719304483},
  urldate = {2023-11-30}
}

@article{schmidt_machinelearningassisted_2023,
  title = {Machine-{{Learning-Assisted Determination}} of the {{Global Zero-Temperature Phase Diagram}} of {{Materials}}},
  author = {Schmidt, Jonathan and Hoffmann, Noah and Wang, Hai-Chen and Borlido, Pedro and Carriço, Pedro J. M. A. and Cerqueira, Tiago F. T. and Botti, Silvana and Marques, Miguel A. L.},
  date = {2023},
  journaltitle = {Advanced Materials},
  volume = {35},
  number = {22},
  pages = {2210788},
  issn = {1521-4095},
  doi = {10.1002/adma.202210788},
  url = {https://onlinelibrary.wiley.com/doi/abs/10.1002/adma.202210788},
  urldate = {2023-10-31}
}

@article{sharma_synthesis_2013,
  title = {Synthesis, Structure, Characterization and Photocatalytic Activity of {{Bi2Zr2O7}} under Solar Radiation},
  author = {Sharma, Vaishali M. and Saha, Dipankar and Madras, Giridhar and Row, T. N. Guru},
  date = {2013},
  journaltitle = {RSC Advances},
  shortjournal = {RSC Adv.},
  volume = {3},
  number = {41},
  pages = {18938},
  issn = {2046-2069},
  doi = {10.1039/c3ra43518k},
  url = {http://xlink.rsc.org/?DOI=c3ra43518k},
  urldate = {2023-11-30}
}

@online{shoghi_molecules_2023,
  title = {From {{Molecules}} to {{Materials}}: {{Pre-training Large Generalizable Models}} for {{Atomic Property Prediction}}},
  shorttitle = {From {{Molecules}} to {{Materials}}},
  author = {Shoghi, Nima and Kolluru, Adeesh and Kitchin, John R. and Ulissi, Zachary W. and Zitnick, C. Lawrence and Wood, Brandon M.},
  date = {2023-10-25},
  eprint = {2310.16802},
  eprinttype = {arxiv},
  eprintclass = {cs},
  doi = {10.48550/arXiv.2310.16802},
  url = {http://arxiv.org/abs/2310.16802},
  urldate = {2023-10-26},
  pubstate = {preprint}
}

@thesis{simon_synthesis_2010,
  title = {The Synthesis and Characterisation of Pyrochlore Frameworks},
  author = {Simon, Charles Francis},
  editora = {Simon, Charles Francis and Weller, Mark},
  editoratype = {collaborator},
  date = {2010-09-30},
  institution = {{University of Southampton}},
  url = {https://eprints.soton.ac.uk/203757/},
  urldate = {2024-01-06},
  pagetotal = {226}
}

@article{soderlind_groundstate_2014,
  title = {Ground-State Properties of Rare-Earth Metals: An Evaluation of Density-Functional Theory},
  shorttitle = {Ground-State Properties of Rare-Earth Metals},
  author = {Söderlind, Per and Turchi, P. E. A. and Landa, A. and Lordi, V.},
  date = {2014-09},
  journaltitle = {Journal of Physics: Condensed Matter},
  shortjournal = {J. Phys.: Condens. Matter},
  volume = {26},
  number = {41},
  pages = {416001},
  publisher = {{IOP Publishing}},
  issn = {0953-8984},
  doi = {10.1088/0953-8984/26/41/416001},
  url = {https://dx.doi.org/10.1088/0953-8984/26/41/416001},
  urldate = {2023-11-21}
}

@article{sorokina_new_1998,
  title = {New Phases in the {{ZrO2}}–{{Bi2O3}} and {{HfO2}}–{{Bi2O3}} Systems},
  author = {Sorokina, S.L. and Sleight, A.W.},
  date = {1998-07},
  journaltitle = {Materials Research Bulletin},
  shortjournal = {Materials Research Bulletin},
  volume = {33},
  number = {7},
  pages = {1077--1081},
  issn = {00255408},
  doi = {10.1016/S0025-5408(98)00076-2},
  url = {https://linkinghub.elsevier.com/retrieve/pii/S0025540898000762},
  urldate = {2023-11-30}
}

@article{steiner_organic_2019,
  title = {Organic Synthesis in a Modular Robotic System Driven by a Chemical Programming Language},
  author = {Steiner, Sebastian and Wolf, Jakob and Glatzel, Stefan and Andreou, Anna and Granda, Jarosław M. and Keenan, Graham and Hinkley, Trevor and Aragon-Camarasa, Gerardo and Kitson, Philip J. and Angelone, Davide and Cronin, Leroy},
  date = {2019-01-11},
  journaltitle = {Science},
  volume = {363},
  number = {6423},
  pages = {eaav2211},
  publisher = {{American Association for the Advancement of Science}},
  doi = {10.1126/science.aav2211},
  url = {https://www.science.org/doi/full/10.1126/science.aav2211},
  urldate = {2024-01-03}
}

@article{sun_thermodynamic_2016,
  title = {The Thermodynamic Scale of Inorganic Crystalline Metastability},
  author = {Sun, Wenhao and Dacek, Stephen T. and Ong, Shyue Ping and Hautier, Geoffroy and Jain, Anubhav and Richards, William D. and Gamst, Anthony C. and Persson, Kristin A. and Ceder, Gerbrand},
  date = {2016},
  journaltitle = {Science Advances},
  volume = {2},
  number = {11},
  pages = {e1600225},
  doi = {10.1126/sciadv.1600225},
  url = {https://www.science.org/doi/abs/10.1126/sciadv.1600225}
}

@article{szymanski_autonomous_2023,
  title = {An Autonomous Laboratory for the Accelerated Synthesis of Novel Materials},
  author = {Szymanski, Nathan J. and Rendy, Bernardus and Fei, Yuxing and Kumar, Rishi E. and He, Tanjin and Milsted, David and McDermott, Matthew J. and Gallant, Max and Cubuk, Ekin Dogus and Merchant, Amil and Kim, Haegyeom and Jain, Anubhav and Bartel, Christopher J. and Persson, Kristin and Zeng, Yan and Ceder, Gerbrand},
  date = {2023-12},
  journaltitle = {Nature},
  volume = {624},
  number = {7990},
  pages = {86--91},
  publisher = {{Nature Publishing Group}},
  issn = {1476-4687},
  doi = {10.1038/s41586-023-06734-w},
  url = {https://www.nature.com/articles/s41586-023-06734-w},
  urldate = {2024-01-03},
  issue = {7990}
}

@article{valant_dielectric_2004,
  title = {Dielectric {{Characteristics}} of {{Bismuth Oxide Solid Solutions}} with a {{Fluorite}}‐{{Like Crystal Structure}}},
  author = {Valant, Matjaz and Suvorov, Danilo},
  date = {2004-06},
  journaltitle = {Journal of the American Ceramic Society},
  shortjournal = {Journal of the American Ceramic Society},
  volume = {87},
  number = {6},
  pages = {1056--1061},
  issn = {0002-7820, 1551-2916},
  doi = {10.1111/j.1551-2916.2004.01056.x},
  url = {https://ceramics.onlinelibrary.wiley.com/doi/10.1111/j.1551-2916.2004.01056.x},
  urldate = {2023-11-30}
}

@article{wang_high_2018,
  title = {High- k {{Gate Dielectrics}} for {{Emerging Flexible}} and {{Stretchable Electronics}}},
  author = {Wang, Binghao and Huang, Wei and Chi, Lifeng and Al-Hashimi, Mohammed and Marks, Tobin J. and Facchetti, Antonio},
  date = {2018-06-13},
  journaltitle = {Chemical Reviews},
  shortjournal = {Chem Rev},
  volume = {118},
  number = {11},
  eprint = {29785854},
  eprinttype = {pmid},
  pages = {5690--5754},
  issn = {1520-6890},
  doi = {10.1021/acs.chemrev.8b00045}
}

@article{wang_predicting_2021,
  title = {Predicting Stable Crystalline Compounds Using Chemical Similarity},
  author = {Wang, Hai-Chen and Botti, Silvana and Marques, Miguel A. L.},
  date = {2021-01-26},
  journaltitle = {npj Computational Materials},
  shortjournal = {npj Comput Mater},
  volume = {7},
  number = {1},
  pages = {1--9},
  publisher = {{Nature Publishing Group}},
  issn = {2057-3960},
  doi = {10.1038/s41524-020-00481-6},
  url = {https://www.nature.com/articles/s41524-020-00481-6},
  urldate = {2022-02-21},
  issue = {1}
}

@article{weiss_photoinduced_2020,
  title = {Photoinduced {{Defect}} and {{Surface Chemistry}} of {{Niobium Tellurium Oxides ANbTeO6}} ({{A}} = {{K}}, {{Rb}}, {{Cs}}) with {{Defect-Pyrochlore Structure}}},
  author = {Weiss, Morten and Wirth, Benedikt and Marschall, Roland},
  date = {2020-06-15},
  journaltitle = {Inorganic Chemistry},
  shortjournal = {Inorg. Chem.},
  volume = {59},
  number = {12},
  pages = {8387--8395},
  issn = {0020-1669, 1520-510X},
  doi = {10.1021/acs.inorgchem.0c00811},
  url = {https://pubs.acs.org/doi/10.1021/acs.inorgchem.0c00811},
  urldate = {2023-11-30}
}

@article{weston_named_2019,
  title = {Named {{Entity Recognition}} and {{Normalization Applied}} to {{Large-Scale Information Extraction}} from the {{Materials Science Literature}}},
  author = {Weston, L. and Tshitoyan, V. and Dagdelen, J. and Kononova, O. and Trewartha, A. and Persson, K. A. and Ceder, G. and Jain, A.},
  date = {2019-09-23},
  journaltitle = {Journal of Chemical Information and Modeling},
  shortjournal = {J. Chem. Inf. Model.},
  volume = {59},
  number = {9},
  pages = {3692--3702},
  publisher = {{American Chemical Society}},
  issn = {1549-9596},
  doi = {10.1021/acs.jcim.9b00470},
  url = {https://doi.org/10.1021/acs.jcim.9b00470},
  urldate = {2020-08-31}
}

@article{wu_preparation_2015,
  title = {Preparation and Photocatalytic Properties of {{Bi2Zr2O7}} Photocatalyst},
  author = {Wu, Deyong and He, Tingting and Xia, Jin and Tan, Yuanbin},
  date = {2015-10},
  journaltitle = {Materials Letters},
  shortjournal = {Materials Letters},
  volume = {156},
  pages = {195--197},
  issn = {0167577X},
  doi = {10.1016/j.matlet.2015.05.107},
  url = {https://linkinghub.elsevier.com/retrieve/pii/S0167577X15300124},
  urldate = {2023-11-30}
}

@article{yan_material_2015,
  title = {Material Descriptors for Predicting Thermoelectric Performance},
  author = {Yan, Jun and Gorai, Prashun and Ortiz, Brenden and Miller, Sam and Barnett, Scott A. and Mason, Thomas and Stevanović, Vladan and Toberer, Eric S.},
  date = {2015-03-05},
  journaltitle = {Energy \& Environmental Science},
  shortjournal = {Energy Environ. Sci.},
  volume = {8},
  number = {3},
  pages = {983--994},
  publisher = {{The Royal Society of Chemistry}},
  issn = {1754-5706},
  doi = {10.1039/C4EE03157A},
  url = {https://pubs.rsc.org/en/content/articlelanding/2015/ee/c4ee03157a},
  urldate = {2020-08-31}
}

@article{yeo_mosfet_2003,
  title = {{{MOSFET}} Gate Leakage Modeling and Selection Guide for Alternative Gate Dielectrics Based on Leakage Considerations},
  author = {Yeo, Yee-Chia and King, Tsu-Jae and Hu, Chenming},
  date = {2003-04},
  journaltitle = {IEEE Transactions on Electron Devices},
  volume = {50},
  number = {4},
  pages = {1027--1035},
  issn = {1557-9646},
  doi = {10.1109/TED.2003.812504},
  eventtitle = {{{IEEE Transactions}} on {{Electron Devices}}}
}

@article{yim_novel_2015,
  title = {Novel High-κ Dielectrics for next-Generation Electronic Devices Screened by Automated Ab Initio Calculations},
  author = {Yim, Kanghoon and Yong, Youn and Lee, Joohee and Lee, Kyuhyun and Nahm, Ho-Hyun and Yoo, Jiho and Lee, Chanhee and Seong Hwang, Cheol and Han, Seungwu},
  date = {2015-06},
  journaltitle = {NPG Asia Materials},
  shortjournal = {NPG Asia Mater},
  volume = {7},
  number = {6},
  pages = {e190-e190},
  publisher = {{Nature Publishing Group}},
  issn = {1884-4057},
  doi = {10.1038/am.2015.57},
  url = {https://www.nature.com/articles/am201557},
  urldate = {2022-03-15},
  issue = {6}
}

@article{zagorac_recent_2019,
  title = {Recent Developments in the {{Inorganic Crystal Structure Database}}: Theoretical Crystal Structure Data and Related Features},
  shorttitle = {Recent Developments in the {{Inorganic Crystal Structure Database}}},
  author = {Zagorac, D. and Müller, H. and Ruehl, S. and Zagorac, J. and Rehme, S.},
  date = {2019-10-01},
  journaltitle = {Journal of Applied Crystallography},
  shortjournal = {J Appl Cryst},
  volume = {52},
  number = {5},
  pages = {918--925},
  publisher = {{International Union of Crystallography}},
  issn = {1600-5767},
  doi = {10.1107/S160057671900997X},
  url = {https://journals.iucr.org/j/issues/2019/05/00/in5024/},
  urldate = {2024-01-05}
}

@article{zhang_finding_2021,
  title = {Finding the {{Next Superhard Material}} through {{Ensemble Learning}}},
  author = {Zhang, Ziyan and Mansouri Tehrani, Aria and Oliynyk, Anton O. and Day, Blake and Brgoch, Jakoah},
  date = {2021},
  journaltitle = {Advanced Materials},
  volume = {33},
  number = {5},
  pages = {2005112},
  issn = {1521-4095},
  doi = {10.1002/adma.202005112},
  url = {https://onlinelibrary.wiley.com/doi/abs/10.1002/adma.202005112},
  urldate = {2023-11-02}
}

@article{zuo_accelerating_2021,
  title = {Accelerating Materials Discovery with {{Bayesian}} Optimization and Graph Deep Learning},
  shorttitle = {{{BOWSR}}},
  author = {Zuo, Yunxing and Qin, Mingde and Chen, Chi and Ye, Weike and Li, Xiangguo and Luo, Jian and Ong, Shyue Ping},
  date = {2021-10-01},
  journaltitle = {Materials Today},
  shortjournal = {Materials Today},
  issn = {1369-7021},
  doi = {10.1016/j.mattod.2021.08.012},
  url = {https://www.sciencedirect.com/science/article/pii/S1369702121002984},
  urldate = {2021-10-21}
}

\clearpage
\appendix

\section*{Supplementary Information}

\section{Related Work}
\label{sec:related-work}

While previous studies have made significant strides in automating high-throughput DFPT to uncover new dielectrics, our work diverges in 3 important regards.
We prefix DFPT with generative and pre-filtering ML which allows us to consider a much larger initial candidate pool as well as venture into uncharted regions of material space in our search for high dielectrics.
Using ML-preselection and biasing the structure generation to crystals similar in chemistry to known high dielectric materials in MP allows us to nonetheless maintain a higher hit rate of materials with high $\fom > 240$ than previous works as shown in \cref{tab:hit-rate-comparison}.
Third, we built a web UI that enabled effective collaboration with experimentalists to select 2 promising candidates which we successfully synthesized and characterized.

To our knowledge, Yim et al. \cite{yim_novel_2015} were the first to develop codes that fully automate ab-initio calculation of band gaps and dielectric permittivities.
They calculated 1800 structures of binary and ternary oxides from the ICSD to generate a dielectric property map which confirmed the inverse correlation between band gap and permittivity for most oxides, with occasional outliers that exhibit both large permittivity despite large band gaps.

\citeauthor*{petousis_benchmarking_2016} \cite{petousis_benchmarking_2016} calculated electronic and ionic dielectric tensors for 88 compounds to test the predictive power of DFPT against experiment for total dielectric constant and refractive index.
While they observed a Mean Average Deviation (MARD) of 16.2\% when using PBE as compared to LDA, they noted that DFPT is less accurate for compounds with complex structural effects or strong anharmonicity.
Their results, however, showed a high Spearman correlation factor of 0.92, demonstrating the utility of DFPT in identifying promising materials by ranking.

The following year, \citeauthor*{petousis_benchmarking_2016}\cite{petousis_high-throughput_2017} extended their previous work by running high-throughput DFPT on 1,056 inorganic compounds.
The resulting database of dielectric tensors was integrated into the Materials Project for public access.
While this greatly improved explorability of the data and likely may have helped expand the search pool for experimentalists seeking synthesis candidates, the scale of the data remained too limited to cover more than a small fraction of compositional and even less of the configurational space of potential high dielectrics.

While the above works resulted in novel and promising candidate materials, they relied exclusively on expensive DFPT calculations, making truly high-throughput screening of hundreds of thousands of materials cost-prohibitive.
Yet they produced a sizeable pool of DFT dielectric properties with which we are now able to train ML models to accelerate and amortize the high cost of DFPT in the search for dielectrics, allowing screening of a much more expansive chemical space.

\section{\texorpdfstring{\BiZrO{}}{Bi2Zr2O7} Synthesis Development and Structure Fitting}
\label{sec:Bi2Zr2O7-synthesis-development}

\BiZrO{} is known and has seen research interest for its use as a photocatalyst \cite{wu_preparation_2015,jayaraman_bridging_2020,luo_new_2018}.
In these reports, the compound has been said to have either a stoichiometric pyrochlore structure (\ch{A2B2O7}) \cite{pandey_metastable_2018,luo_new_2018,liu_bi2zr2o7_2018,luo_synthesis_2019,jayaraman_bridging_2020,kurlla_greenengineered_2023} or the structurally related defect fluorite structure \cite{sorokina_new_1998,sharma_synthesis_2013,wu_preparation_2015,rajashekharaiah_nuv_2019,feng_unraveling_2021,}.
Our results show that a pyrochlore could not be isolated without additional \ch{Bi2O3} or \ch{ZrO2} impurities due to the metastable nature of this compound.
Though the pyrochlore and fluorite structures yield similar XRD patterns, with the most intense peaks located in the same positions, the absence of the (111) peak at $Q = 1.01\A^{-1}$ favors assignment of the fluorite structure, \cref{fig:exp-rietveld-Bi2Zr2O7-Fm3m}.

The XRD data was fit using Rietveld refinement.
Attempts were made to fit the data with a pyrochlore structure.
When using both a standard pyrochlore model and models with oxygen and \ch{Bi3+}, displacive disorder produces calculated patterns that fail to fit the data properly.
Intensity mismatch is observed for low-angle pyrochlore peaks, specifically the (111) reflection.
No amount of disorder was sufficient to reduce the intensity of this peak in the model to noise levels in the data, further confirming that this compound does not crystallize as a pyrochlore.

Using a defect fluorite structure (\ch{Bi_{0.5}Zr_{0.5}O_{1.75}}, \cref{fig:Bi2Zr2O7-whole-cell}) results in rapid model-to-data convergence with a good visual fit, \cref{fig:exp-rietveld-Bi2Zr2O7-Fm3m}.
The resultant model shows atomic displacement parameters of 3.28(17) $\A^2$ for the cations and 8.4(4) $\A^2$ for the oxygen, which are large.
Large atomic displacement parameters are commonly found in disordered compounds, and experimentally observable in the form of broad diffraction peaks, compared to \cref{fig:exp-rietveld-CsTaTeO6-Fd3m}.
Attempts to account for the disorder in this compound in our structural model were not successful.
Splitting the position of Zr and Bi to account for chemical displacements off their position in the center of the cubic polyhedra resulted in the cations refining back to their undisplaced positions.
The same process was used for the oxygen positions but resulted in much larger atomic displacement parameters, leading us to discount this distortion.
The occupancies of sites were also refined, resulting in the cations maintaining a 1:1 ratio, within error.
This allowed us to conclude that a simple defect fluorite structure is the most sensible model.
This final model can be seen in \cref{fig:Bi2Zr2O7-whole-cell} and an isolated \ch{Zr/BiO8} polyhedra can be seen in \cref{fig:Bi2Zr2O7-polyhedra}.
This model produced sensible metal-oxygen bond lengths of 2.3160(5) \A, expectedly longer than \ch{ZrO2} bond lengths of 2.25 \A.

\section{\texorpdfstring{\CsTaTeO{}}{CsTaTeO6} Synthesis Development and Structure Fitting}
\label{sec:CsTaTeO6-synthesis-development}

The targeted \CsTaTeO{} pyrochlore compound was initially investigated as it both met the figure of merit criterion and had not been reported previously in the ICSD or MP.
However, we did find mention of this compound and its crystallographic analysis in \cite{simon_synthesis_2010} after completing synthesis and characterization.
Moreover, a related pyrochlore with composition \ch{CsNbTeO6} had been reported in \cite{fukina_structure_2021,weiss_photoinduced_2020} from which we extracted initial synthesis parameters.
With minor modifications of the synthetic procedure, the new compound was isolated in high purity, with only a 4.20 wt\% \ch{Ta2O5} impurity.

\Cref{fig:exp-rietveld-CsTaTeO6-Fd3m} shows XRD data of the final \CsTaTeO{} product.
This pattern indexes readily to the symmetry and lattice parameters of a cubic pyrochlore (\cref{fig:CsTaTeO6-whole-cell}), consistent with both the \ch{Nb^5+}-based analog and the computational predictions.
Rietveld refinements were initiated using parameters taken from Pawley fitting and readily converged to a pyrochlore structural model.
The structural model was taken from the refinement of \ch{CsNbTeO6} which places \ch{Cs} on the larger site 8b (\cref{fig:CsTaTeO6-a-site}) site and the \ch{Ta^5+} and \ch{Te^6+} in equimolar amounts on the 16c site (\cref{fig:CsTaTeO6-b-site}).
This formulation is that of a defect pyrochlore (\ch{AB2O6}), which is distinct from the traditional \ch{A2B2O7} pyrochlore structure.
Relative to a traditional pyrochlore, this structure has both cation and anion vacancies, while maintaining the same anion packing and \ch{BO6} connectivity.
After refining all parameters simultaneously, a good visual fit to the data is obtained with sensible atomic positions, sensible atomic displacement parameters (0.087(15) – 0.78(2) $\A^2$), a lattice parameter (10.29894(5) \A) close to that of the \ch{Nb^5+} analog (10.288 \A), and a fit quality parameter (Rwp = 8.095\%) approaching that of the minimum set by the Pawley Fit (Rwp = 7.399\%) \cite{galati_cation_2008}.

Due to the defect nature of this pyrochlore formulation, the \ch{Cs+} (A-site) adopts an octahedral polyhedral environment (\cref{fig:CsTaTeO6-a-site}) with six equal bond lengths of 3.183(6) \A, instead of the cubic (AO8) environment found in stoichiometric \ch{A2B2O7} pyrochlores.
The observed bond lengths are consistent with \ch{AO6} polyhedral environments seen in other \ch{Cs+} pyrochlores such as \ch{CsNbTeO6} or \ch{CsMoTeO6} which range from 3.180 – 3.421 \A \cite{galati_cation_2008,fukina_crystal_2019}.
The \ch{Ta^5+} and \ch{Te6+} occupy the smaller octahedral environment (\cref{fig:CsTaTeO6-b-site}) found in traditional and defect pyrochlores.
This environment generates bond lengths of 1.9430(18) \A, again falling within the expected range of related materials such as the \ch{Nb^5+} and \ch{Mo^5+} analogs previously mentioned, compounds that range from 1.941 – 2.013 \A.
This consistency of the structural environments found in \CsTaTeO{} with similar chemistries further validates the quality of our model.

\section{Tradeoffs in Dielectric Materials for Computing Applications}

As indicated by the shaded regions in \cref{fig:diel-total-vs-bandgap-mp}, while ideal dielectric materials all push into the top right of this plot, different applications have different requirements.
Materials for flash storage require especially large band gaps to minimize leakage current and maintain polarization over extended periods.
CPU gate dielectrics trade off lower band gaps in exchange for increased permittivity which lowers the gate voltage required to achieve polarization and hence decreases power consumption.
For random access memory (RAM) applications, increased leakage current resulting from a lower band gap is acceptable since RAM is memory-refreshed hundreds of times a second (stored data is read and immediately rewritten unmodified to preserve integrity to avoid polarization sapping over time).
Instead, optimal RAM performance relies on exceptionally high permittivity so that each repolarization costs minimal energy.
Our goal is to discover materials in any of these regions beyond the green isoline ($\fom = 240$).
\cref{fig:diel-parts-vs-bandgap-mp} shows that the principal contributions to the permittivity are due to the ionic permittivity of the materials rather than their electronic permittivity.

\begin{figure}
    \centering
    \includegraphics[width=0.68\columnwidth]{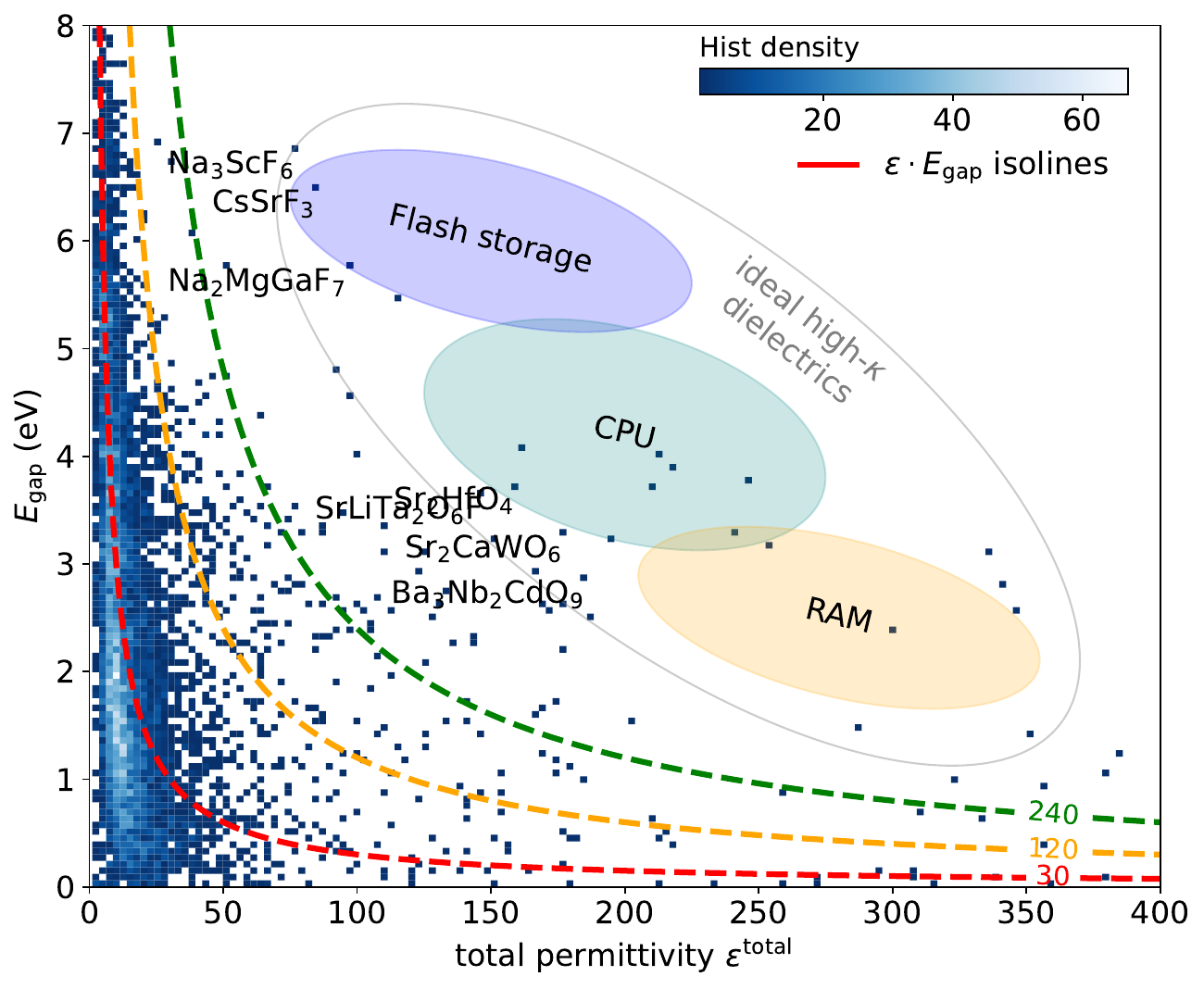}
    \caption{2d histogram showing the $1/x$ relationship between band gap and dielectric constant for 7.2k MP materials. The dashed isolines represent levels of constant figure of merit ($\epstot \cdot \egap$). The colored ellipses highlight the optimal trade-offs between band gap and permittivity for specific device applications. See \cref{fig:diel-parts-vs-bandgap-mp} for the same plot split by electronic and ionic contributions to the permittivity.}
    \label{fig:diel-total-vs-bandgap-mp}
\end{figure}

\begin{figure}[ht!]
    \centering
    \includegraphics[width=\linewidth]{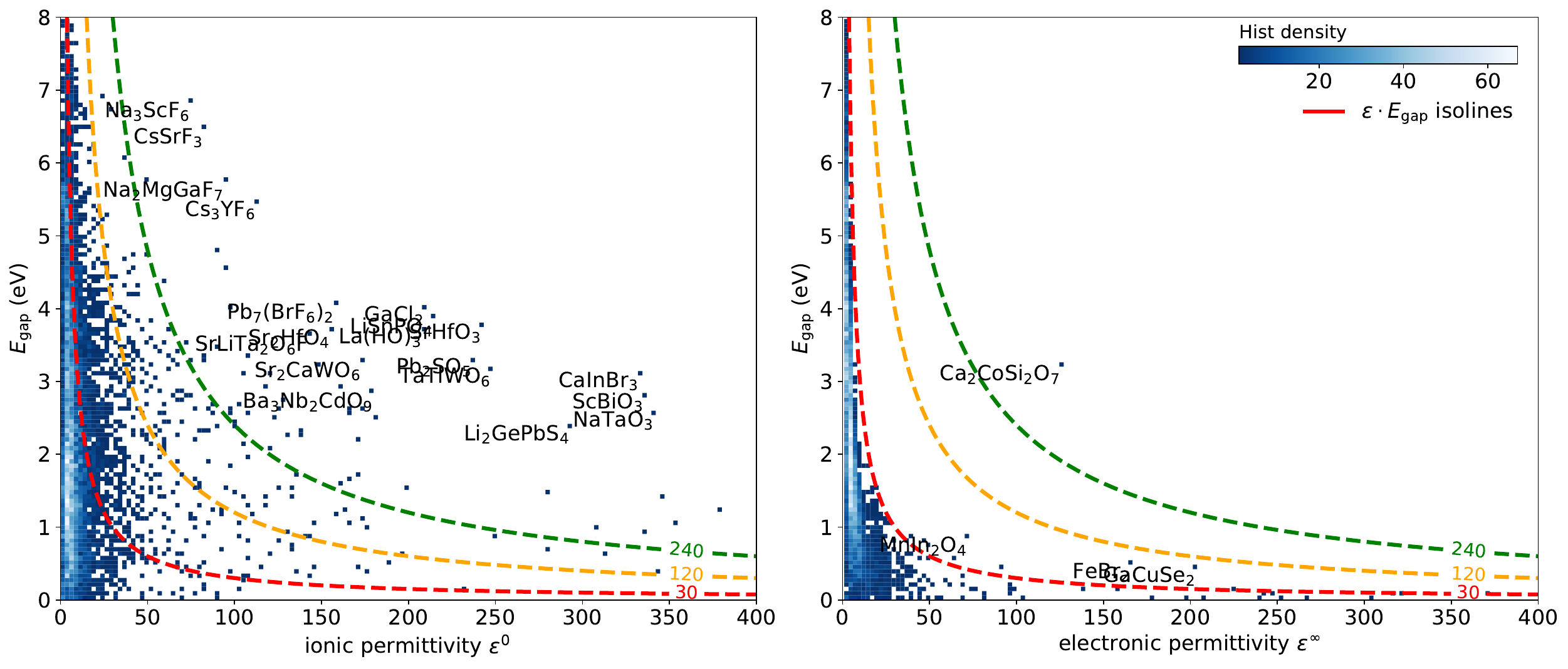}
    \caption{
        Ionic (left) and electronic (right) parts of the dielectric constant.
        Ionic permittivity makes a much larger contribution than the total compared to electronic permittivity and is also much more likely to break the 1/x relationship with band gap.
    }
    \label{fig:diel-parts-vs-bandgap-mp}
\end{figure}

\clearpage
\section{DFPT Validation}
\label{sec:dfpt-validation}

To validate our DFPT results, \cref{fig:exp-vs-us-vs-petousis-diel-total} compares our ab-initio results generated using the \texttt{wf\_dielectric\_constant} workflow in \texttt{atomate} \cite{mathew_atomate_2017} against available experimental dielectric constants collected in \cite{petousis_benchmarking_2016,petousis_high-throughput_2017}.
While we achieve better agreement with experiment than Petousis as indicated by the lower MAE of 16.5 (vs 20.4) and higher $R^2$ of 0.41 (vs 0.0), and similar performance to MP (MAE = 14.9, $R^2 = 0.12$), we incur a slightly larger fraction of outliers than either of them at 14\% (vs 9\% and 10\%, respectively). We define outliers as points with absolute relative deviation greater than $\pm 50\%$ relative to experiment.
The reason we nonetheless achieve higher $R^2$ is due to the lack of extreme outliers; we have more but they are less severe.
This is advantageous in high-throughput settings where the goal is to guide experiment.
Even rare cases of extreme outliers will show up given sufficient throughput and extreme permittivity overpredictions are more likely to result in wasted experimental effort.

We note that while the data in MP was generated with the same \texttt{atomate} workflow
as designed and benchmarked by \citeauthor*{petousis_benchmarking_2016}\cite{petousis_benchmarking_2016},
our data is expected to deviate from MP/Petousis due to our departure in choice of VASP parameters described in \cref{sec:dfpt-to-dielectric}, most notably the use of \texttt{PBE\_54} POTCARs, increased $k$-point density of 3000 points per atom, increased \texttt{ENCUT} = \SI{700}{eV} plane wave energy cutoff and decreased \texttt{EDIFF} = \SI{e-7}{eV} SCF convergence criterion.
All of the above, though most notably the newer pseudopotentials may explain the less extreme outliers with respect to experiment.
Overall, the variations with respect to MP/Petousis are within reason for run-to-run variability using slightly modified settings.

\begin{figure}[htbp!]
    \centering
    \includegraphics[width=\linewidth]{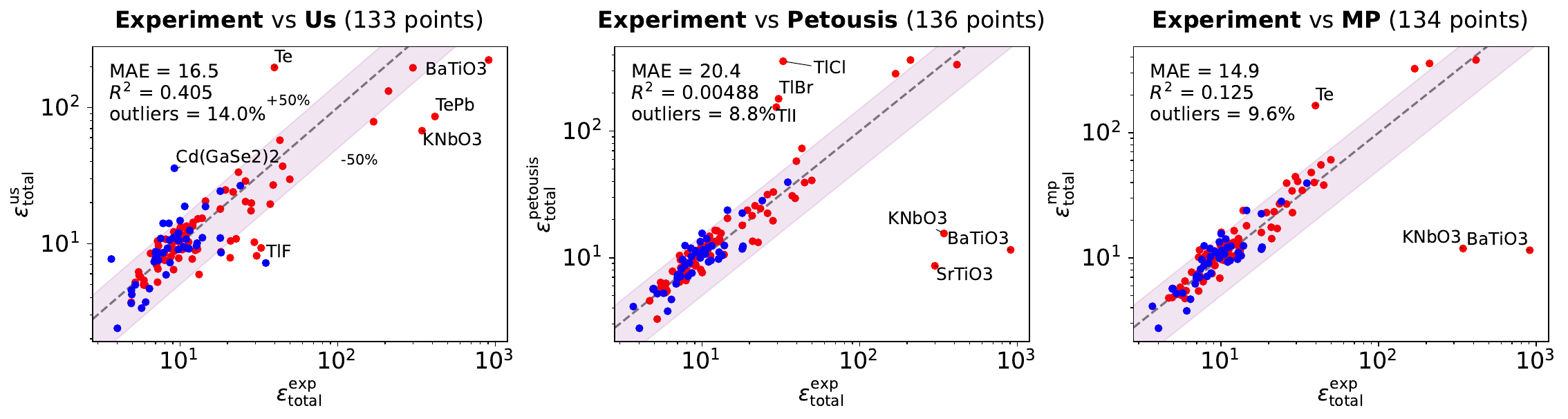}
    \caption{
        Comparison of experimental and DFPT-computed values for total permittivity $\epstot$.
        Our data shows lower MAE and higher $R^2$ but more outliers (defined as points with $>50\%$ error) compared to Petousis et al.
        Comparing our DFPT dielectric constants with experimental values, we achieve an MAE of 16.5 and $R^2$ of 0.4 while MP results attain a slightly lower MAE of 14.9 and $R^2$ of 0.125.
        A CSV file with the plotted experimental data is available on \href{hhttps://github.com/janosh/dielectrics/blob/68839f9d8/data/others/petousis/exp-petousis.csv}{GitHub}.
    }
    \label{fig:exp-vs-us-vs-petousis-diel-total}
\end{figure}

\clearpage

\section{Exploratory Data Analysis}
\label{sec:exploratory-data-analysis}

\begin{figure}[htbp!]
    \centering
    \begin{subfigure}[b]{0.495\linewidth}
        \includegraphics[width=\linewidth]{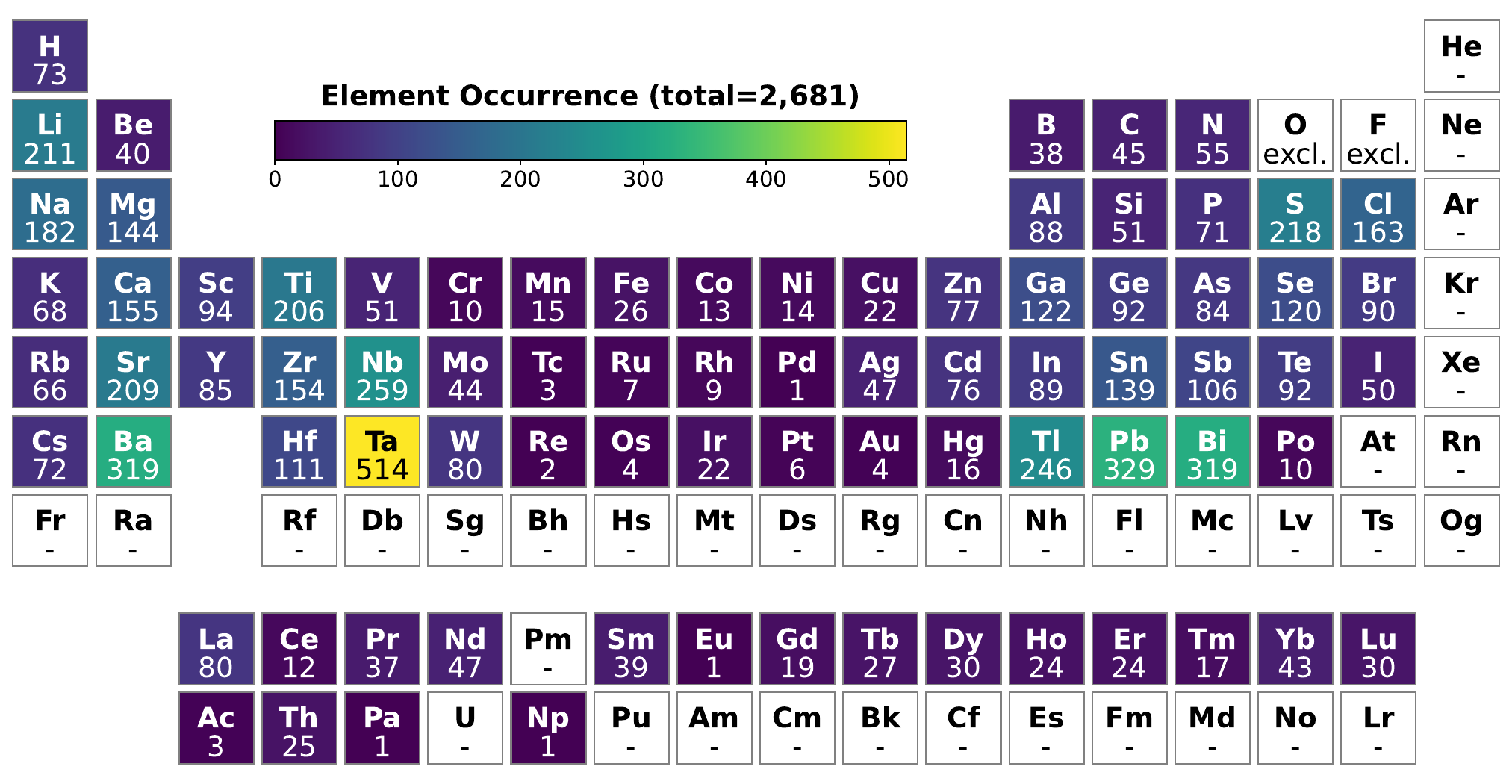}
        \caption{Element occurrences}
        \label{fig:ptable-elem-counts-us}
    \end{subfigure}
    \begin{subfigure}[b]{0.495\linewidth}
        \includegraphics[width=\linewidth]{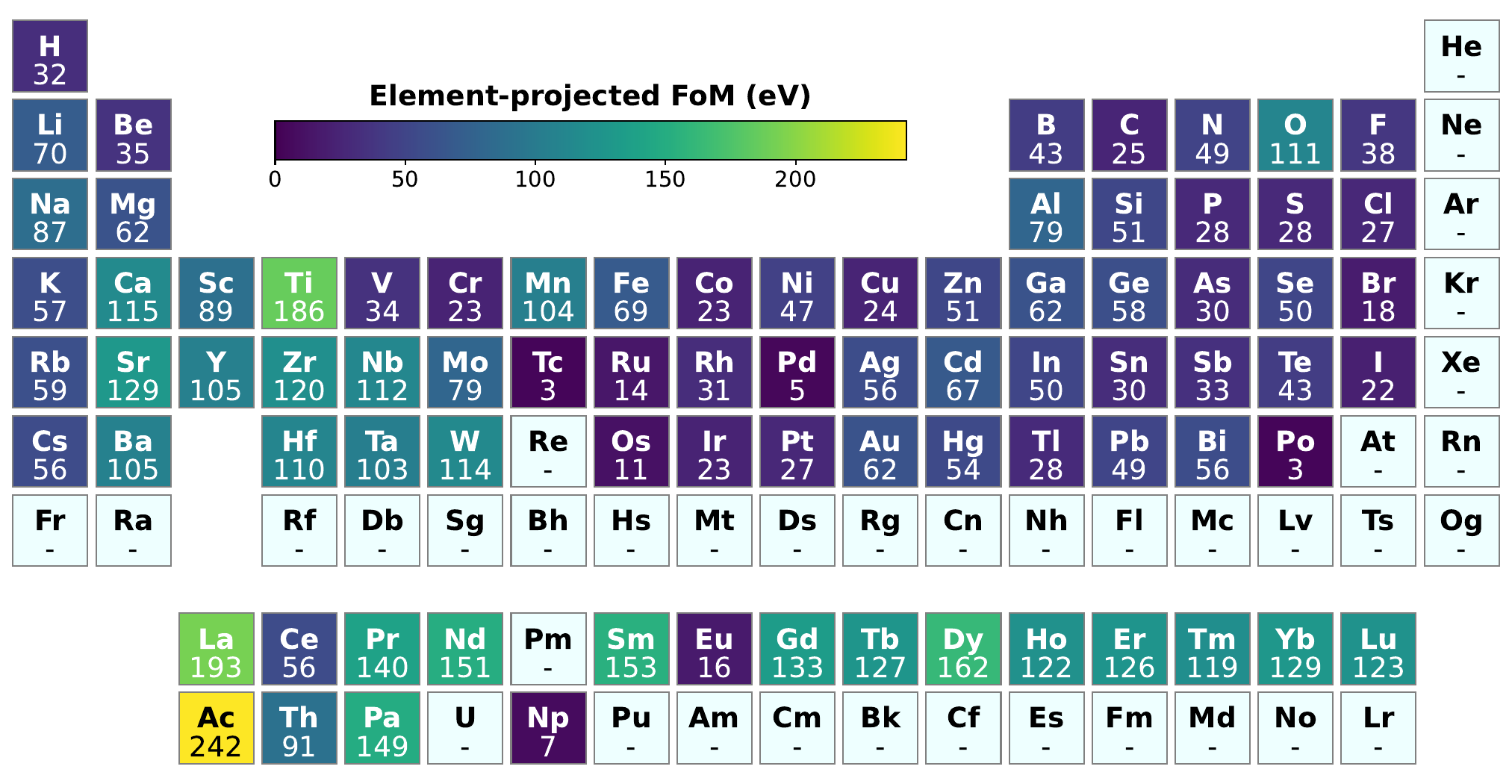}
        \caption{$\fom = \epstot \cdot \egap$}
        \label{fig:ptable-per-elem-fom-pbe}
    \end{subfigure}
    \begin{subfigure}[b]{0.495\linewidth}
        \includegraphics[width=\linewidth]{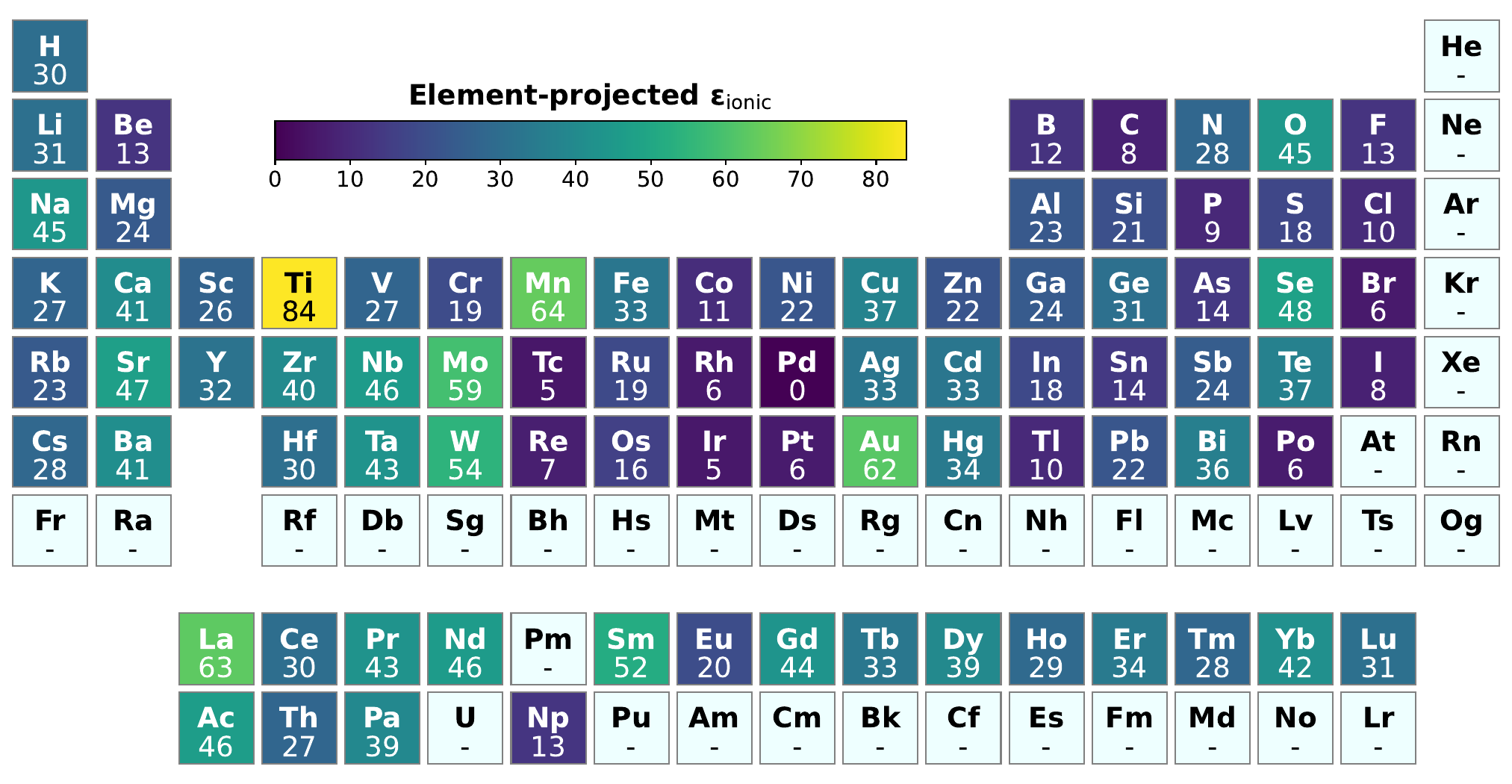}
        \caption{Ionic permittivity $\epsilon_0$}
        \label{fig:ptable-per-elem-diel-ionic-pbe}
    \end{subfigure}
    \begin{subfigure}[b]{0.495\linewidth}
        \includegraphics[width=\linewidth]{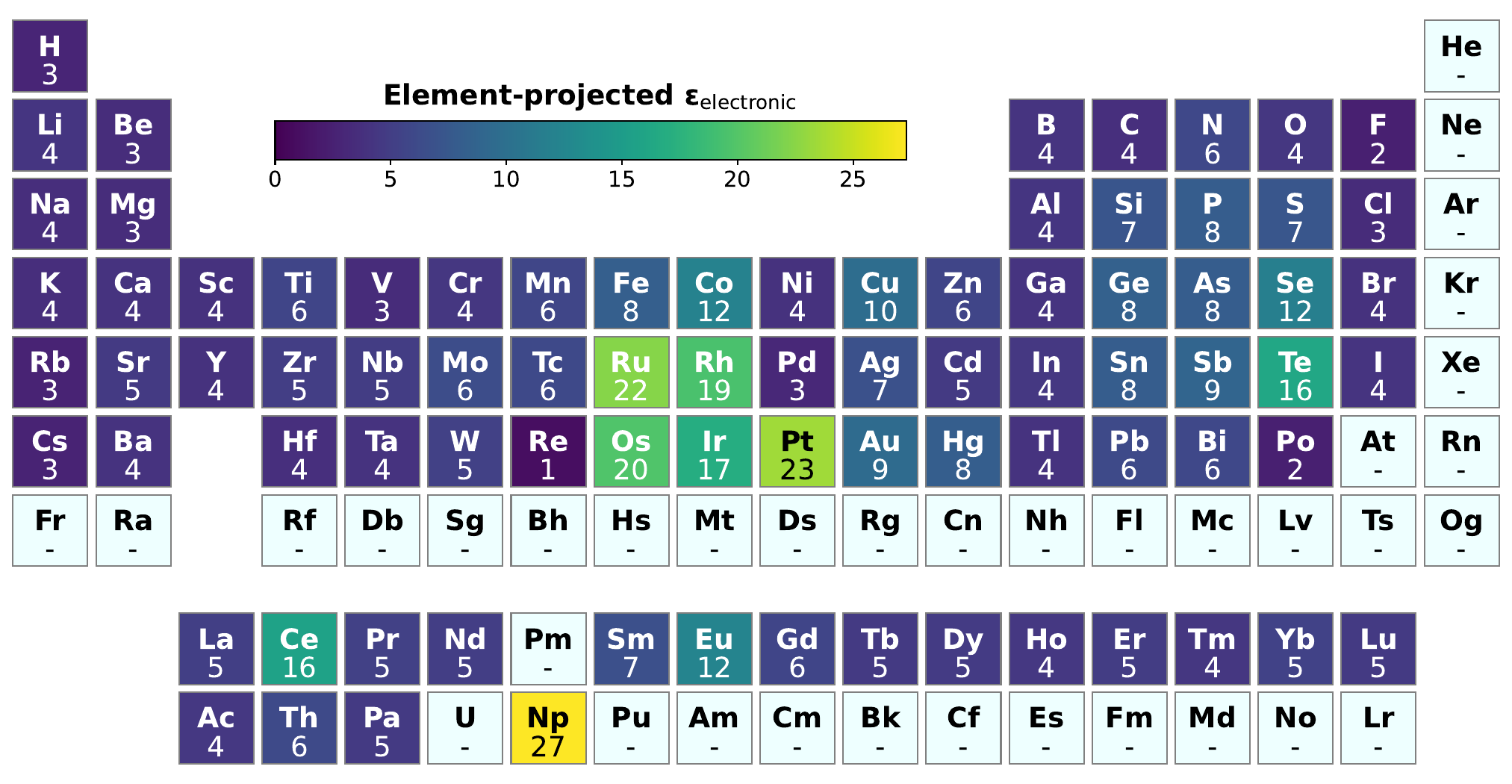}
        \caption{Electronic permittivity $\epsilon_\infty$}
        \label{fig:ptable-per-elem-diel-elec-pbe}
    \end{subfigure}
    \caption{
        \subref*{fig:ptable-elem-counts-us}) Element occurrence counts, i.e. the number of structures among \num{2681} DFPT results containing a given element.
        \subref*{fig:ptable-per-elem-fom-pbe}) Figure of merit $\fom = \epstot \cdot \egap$ projected onto elements by composition and averaged over all \num{2681} structures.
        E.g. a \ch{Fe2O3} with a $\fom$ of 100 would contribute a sample of 40 to the mean heatmap value of \ch{Fe} and 60 to \ch{O}.
        \subref*{fig:ptable-per-elem-diel-ionic-pbe}) same as \subref*{fig:ptable-per-elem-fom-pbe} but for ionic permittivity $\epsilon_0$.
        \subref*{fig:ptable-per-elem-diel-elec-pbe}) same as \subref*{fig:ptable-per-elem-fom-pbe} but for electronic permittivity $\epsilon_\infty$.
        We filtered out 10 untrustworthy calculations with electronic permittivity $\epsilon_\infty > 100$.
    }
    \label{fig:ptable-per-elem}
\end{figure}

\Cref{fig:ptable-per-elem} shows the distribution of elements in our DFPT dataset (\subref*{fig:ptable-elem-counts-us}) and their element-projected figure of merit $\fom$ (\subref*{fig:ptable-per-elem-fom-pbe}) and electronic (\subref*{fig:ptable-per-elem-diel-ionic-pbe}) and ionic (\subref*{fig:ptable-per-elem-diel-elec-pbe}) permittivities $\epsilon_\infty$ and $\epsilon_0$.
The most prevalent elements in our dataset are \ch{Ta} (514), \ch{Pb} (329), \ch{Bi} (319), \ch{Ba} (319), \ch{Nb} (259) where the number in parentheses is the number of structures containing that element.
Our data recovers well-known trends for elements that tend to be present in high dielectrics.
In particular, \cref{fig:ptable-per-elem-diel-ionic-pbe} shows titanium has the highest ionic permittivity when averaged over all \ch{Ti}-containing structures in our dataset.
This matches the prevalence of high dielectric alkaline earth metal titanates such as the perovskites \ch{BaTiO3} and \ch{SrTiO3}.
\Cref{fig:ptable-per-elem-diel-elec-pbe} reveals that late transition metals like \ch{Ru}, \ch{Rh}, \ch{Os}, \ch{Ir} and \ch{Pt} tend to yield the highest observed electronic permittivities.

\Cref{tab:table-fom-pbe-gt-350} lists all DFPT results in our dataset with $\fom > 350$ sorted by $\fom$.
The highest-$\fom$ materials are almost exclusively oxides with only two fluorides and one selenide in the mix (\ch{AcF3}, \ch{LiY2F7} and \ch{Sm2CdSe4}).
Some of the top materials, unfortunately, contain toxic or rare elements (e.g. \ch{Cd}, \ch{Nd}, \ch{Dy}) which are undesirable for environmental, economic and lab-safety/logistic reasons.
Others contain lanthanides and actinides, f-block elements which DFT is known to struggle with due to strong electron correlation effects in the atomic-like $4f$ orbitals near the valence band \cite{soderlind_groundstate_2014}.
Both are strong arguments against attempting experimental synthesis, explaining why we did not simply select the top materials in this list.

\begin{table}[htbp!]
    \centering
    \includegraphics[height=0.75\paperheight, width=\linewidth, keepaspectratio]{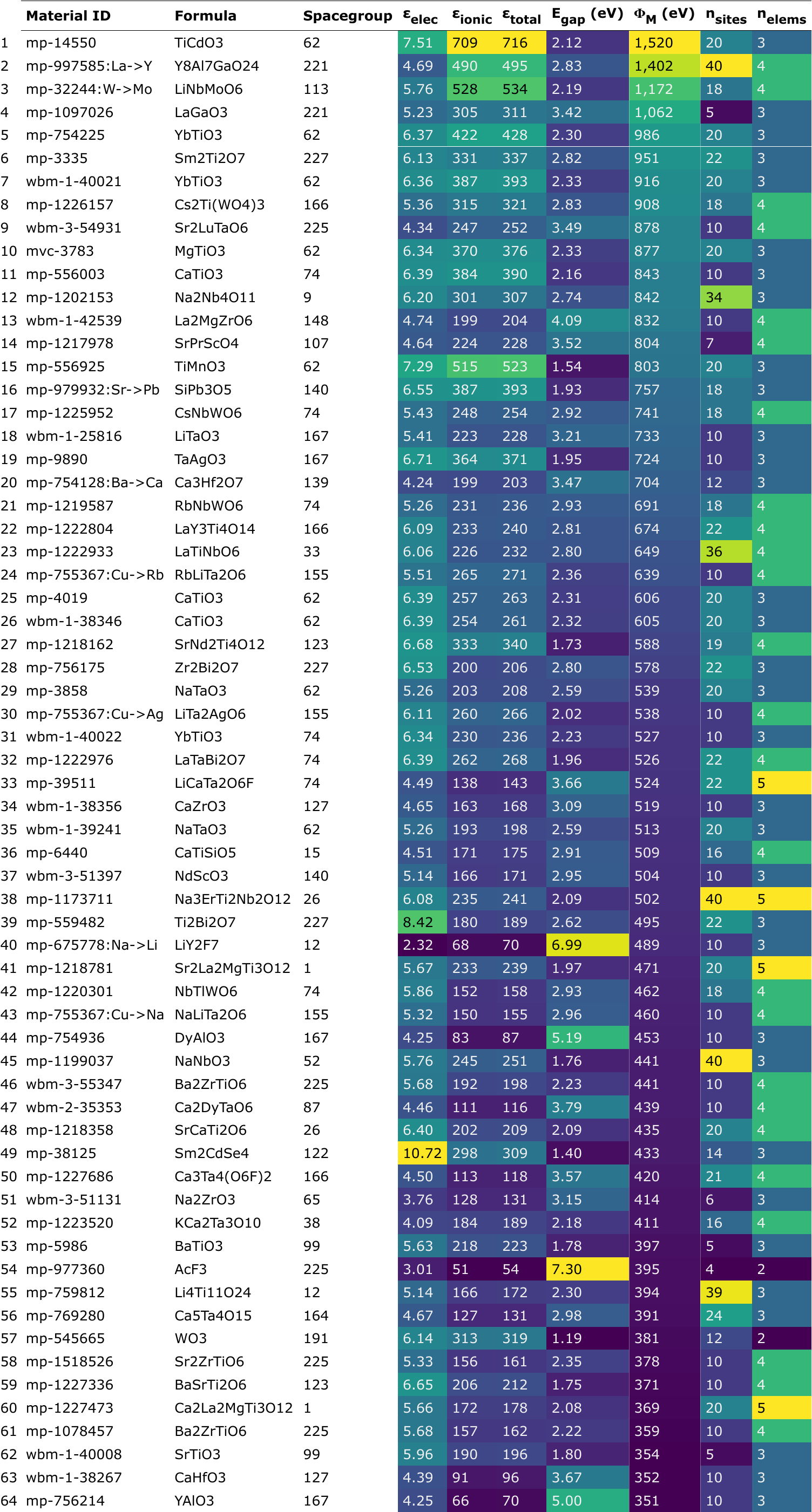}
    \caption{
        Materials with DFPT-computed $\fom > 350$, sorted by $\fom$.
        While these are the highest-reward materials from a purely computational standpoint, synthesis of these high-arity compounds is made challenging by the proliferation of competing in higher dimensional chemical spaces.
        Many of the listed compounds therefore have a risk-reward profile of lower appeal than other materials in our dataset with lower-predicted $\fom$.
        A CSV file of this table is available on \href{https://github.com/janosh/dielectrics/blob/b6f46410e8fcafd727006956380867d476e32177/data/our-data-with-fom-pbe-gt-350.csv}{GitHub}.
    }
    \label{tab:table-fom-pbe-gt-350}
\end{table}

\clearpage
\section{Band Gap Prediction}
\label{sec:band-gap-model}

\begin{figure}[ht!]
    \centering
    \begin{subfigure}[b]{0.495\linewidth}
        \includegraphics[width=\linewidth]{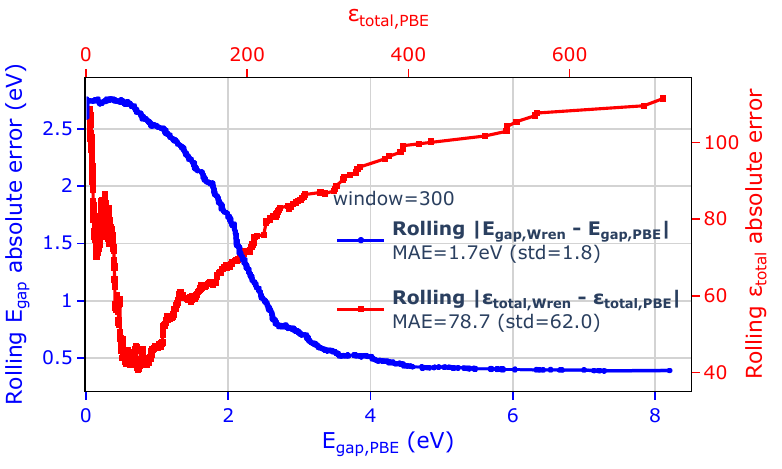}
        \caption{RAE as a function of PBE band gap}
        \label{fig:rolling-bandgap+diel-error-pbe-as-x}
    \end{subfigure}
    \hfil
    \begin{subfigure}[b]{0.495\linewidth}
        \includegraphics[width=\linewidth]{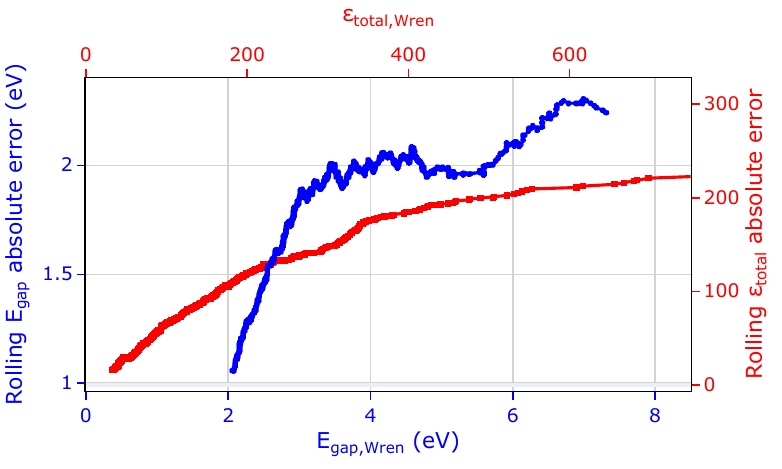}
        \caption{RAE as a function of Wren band gap}
        \label{fig:rolling-bandgap+diel-error-wren-as-x}
    \end{subfigure}
    \caption{
        Rolling absolute error (RAE) of Wren band gap and dielectric constant predictions relative to DFPT.
    }
    \label{fig:rolling-bandgap+diel-error}
\end{figure}

In this screening campaign, we emphasize that substantial challenges remain concerning band gap prediction, due in part to a metal-heavy dataset imbalance and in part to band gaps being an inherently non-local property of the electronic rather than ionic structure.
This makes the prediction problem poorly, if not ill-defined for a coarse-grained structural model with no concept of electronic degrees of freedom.
We describe some attempts to mitigate this issue that achieved limited success but ultimately consider ML band gap prediction a high-impact but unsolved problem (see discussion \cref{sec:discussion}).

Our ensemble of 10 Wren band gap models trained with L1 loss achieved a deceivingly low $\text{MAE} = \SI{0.151}{eV}$ and high coefficient of determination $R^2 = 0.969$.
This is largely due to the aforementioned dataset imbalance.
\num{243095} / \num{319601} = 76.1\% of the combined MP + WBM dataset are PBE metals.
Not wanting to discard 3/4 of our training data, we attempted naive equal loss weighting across all samples as well as increased loss weighting of non-metals.
Finally, we tried prepending a metal-nonmetal classifier to our band gap regressor to only predict the band gap for materials classified as non-metals.
While the latter slightly decreased the false-positive rate, neither managed to significantly improve the overall performance of our band gap model nor fix this main failure mode in our discovery pipeline of metals classified as insulators/semiconductors.
Many of the generated elemental substitution structures we predicted to have sizable band gaps turned out to be PBE metals.
More recent efforts in training foundation models on giant datasets and then fine-tuning on smaller cognate datasets \cite{shoghi_molecules_2023} have achieved impressive sub-\SI{100}{eV/atom} band gap MAEs and may be able to overcome this issue.

\Cref{fig:rolling-bandgap+diel-error} plots the rolling absolute error of our Wren band gap and dielectric constant ensembles with respect to DFPT using a variable window size of 300 samples.
In \cref{fig:rolling-bandgap+diel-error-pbe-as-x}, the bottom x-axis spans the range of PBE-computed band gaps for which we also have Wren predictions.
Similarly, the top x-axis spans the range of DFPT-computed dielectric constants for which we also have Wren predictions.
In \cref{fig:rolling-bandgap+diel-error-wren-as-x}, we swap the x-axis values to be PBE instead. That is Wren band gap predictions on the bottom x-axis and Wren dielectric constants on the top x-axis.
The y-axis is identical in both subplots: the rolling Wren-vs-DFPT absolute error for band gaps on the left and dielectric constants on the right.

\Cref{fig:rolling-bandgap+diel-error-pbe-as-x} reveals that the error in dielectric constant shows a pronounced dip at intermediate ranges from about 40 to 120.
This supports our initial argument for choosing dielectrics as the target material class for this discovery campaign.
We hypothesized that by optimizing the trade-off between two opposing material properties, we can operate both the dielectric and band gap models in regions of good training support where ML models are most reliable and still discover materials with high $\fom$.
In the case of the band gap model, this argument is less supported by the data.
While the error in band gap prediction indeed drops significantly in our target region of $\egap > \SI{2}{eV}$, the error does not increase again for extreme values but stays low even for outlier points beyond \SI{5}{eV}.
However, small errors on large band gaps do not negatively affect the chances of dielectric materials discovery and so are not in conflict with our objective.
The issue with the band gap model is that its error for small band gaps is $> \SI{2}{eV}$ and therefore large enough to predict metals as insulators, thereby introducing false positives into our discovery pipeline.

\Cref{fig:rolling-bandgap+diel-error-wren-as-x} reveals that our workflow suffered from a negative feedback loop in that we purposely selected materials with large band gaps according to Wren which drew the bulk of our selection towards the lower end of the blue line. This line ends at a minimum band gap of \SI{2}{eV}, indicating that no smaller Wren band gaps made it into our DFPT validation set.
However, this is precisely the region where model error and its prediction are almost equal, resulting in a large number of false positive insulator predictions that turned out to be PBE metals.

\end{document}